# Critical phenomena in the temperature-pressure-crowding phase diagram of a protein

Andrei G. Gasic,[1,2,*] Mayank M. Boob,[3,*] Maxim B. Prigozhin,[4,†] Dirar Homouz,[1,2,5]
Caleb M. Daugherty,[1,2] Martin Gruebele,[3,4,6,‡] and Margaret S. Cheung[1,2,§]

[1]*University of Houston, Department of Physics, Houston, Texas, 77204, United States.*

[2]*Center for Theoretical Biological Physics, Rice University, 77005, United States.*

[3]*Center for Biophysics and Quantitative Biology, University of Illinois at Urbana-Champaign, Champaign, IL 61801, United States.*

[4]*Department of Chemistry, University of Illinois at Urbana-Champaign, Champaign, IL, 61801, United States.*

[5]*Khalifa University of Science and Technology, Department of Physics, P.O. Box 127788, Abu Dhabi, United Arab Emirates.*

[6]*Department of Physics and Beckman Institute for Advanced Science and Technology, University of Illinois at Urbana-Champaign, Champaign, IL 61801, United States.*

[*]These authors contributed equally to this work.
[†]Present address: Department of Physics, James H. Clark Center, Stanford University, Stanford, CA, 94305, United States.
[‡]mgruebel@illinois.edu
[§]mscheung@uh.edu

# Abstract

In the cell, proteins fold and perform complex functions through global structural rearrangements. Function requires a protein to be at the brink of stability to be susceptible to small environmental fluctuations, yet stable enough to maintain structural integrity. These apparently conflicting behaviors are exhibited by systems near a critical point, where distinct phases merge—a concept beyond previous studies indicating proteins have a well-defined folded/unfolded phase boundary in the pressure-temperature plane. Here, by modeling the protein phosphoglycerate kinase (PGK) on the temperature ($T$), pressure ($P$), and crowding volume-fraction ($\phi$) phase diagram, we demonstrate a critical transition where phases merge, and PGK exhibits large structural fluctuations. Above the critical temperature ($T_c$), the difference between the intermediate and unfolded phases disappears. When $\phi$ increases, the $T_c$ moves to a lower $T$. We verify the calculations with experiments mapping the $T$-$P$-$\phi$ space, which likewise reveal a critical point at 305 K and 170 MPa that moves to a lower $T$ as $\phi$ increases. Crowding places PGK near a critical line in its natural parameter space, where large conformational changes can occur without costly free energy barriers. Specific structures are proposed for each phase based on simulation.

## I. INTRODUCTION

Complex processes in nature often arise at an order-disorder transition [1–4]. In proteins, this complexity arises from an almost perfect compensation of entropy by enthalpy: molecular interactions that create structural integrity are on the same scale as thermal fluctuations from the environment. The resulting marginal stability of proteins suggests that they could behave like fluids near a critical point [5] – their structures fluctuate considerably subject to small perturbations without overcoming a large activation barrier.

The concept of first-order and critical phase transitions does not rigorously apply to nano-objects such as proteins; nevertheless, it is a useful one to classify folding transitions. For example, folding of small model proteins has been described as an abrupt, cooperative transition between the folded and unfolded phase for some proteins (the below-critical point scenario), or as a gradual barrier-less 'downhill' transition for other proteins (the above-critical point scenario) [6]. Even though critical behavior of proteins has been previously hinted [7–9], there has not been a direct observation of a critical point where one of these abrupt transitions simply disappears at $T_c$ and $P_c$. In larger proteins, such as phosphoglycerate kinase (PGK) ( [10] in section S1), the situation can get even more complex: different parts or 'domains' of a large protein are more likely to be able to undergo separate order-disorder events [11], delicately poised between folded and partially unfolded structures to carry out their functions [12].

Proteins must fold and function while crowded by surrounding macromolecules [13], which perturb the structure of the proteins at physiological conditions in the cell. The volume exclusion from macromolecules [14], which places shape and size (or co-volume) [15,16] constraints on the conformational space [Fig. 1(a)], complicates protein folding and dynamics in living cells [17]. How the competing properties of a protein arise – being both stable yet dynamically sensitive to its environment – is mostly unknown; however, we show that the crowded environment provides a unique solution by placing PGK near a critical regime.

We use pressure $P$, temperature $T$, and crowder-excluded volume fraction $\phi$, to map PGK's folding energy landscape [12,18] and its critical regime on the $T$-$P$-$\phi$ phase diagram. Temperature can induce heat unfolding by favoring states of high conformational entropy, or cold denaturing by favoring reduced solvent entropy when hydrating core amino acids in the protein [19,20]. Since folded proteins contain heterogeneously distributed small, dry cavities due to imperfect packing of their quasi-fractal topology [21–23], high pressure also induces unfolding by introducing water molecules (as small granular particles) into the cavities in protein structures, leading to a reduced overall solvent-accessible volume of the unfolded protein [24]. Finally, in the presence of high crowding (large excluded volume fraction $\phi$), compact desolvated (crystal) states are favored over less compact solvated (unfolded) states [12].

To investigate the opposing impact of macromolecular volume exclusion and solvation water on protein conformation, we utilized a minimalist protein model (see Appendix A and [10] in section S2.2) that incorporates the free energy cost of expelling a water molecule between a pair of residues in a contact termed the desolvation potential [Fig. 1(b)] [25]. This potential has a barrier that separates two minima that account for a native contact and a water-mediated contact. As pressure increases, the desolvation barrier increases and the free energy gap between the two minima tilts to favor the water-mediated contact, leading to an unfolding of a protein, capturing the main feature of pressure denaturation. Despite the model's simplicity without all the detailed chemistry in a residue [26], this desolvation model predicts a folding mechanism involving water expulsion from the hydrophobic core, which has been observed by all-atomistic molecular dynamics [27] and validated by experiments in which the volume or polarity of amino acids is changed by mutation [28]. We previously employed a similar model without desolvation potential to investigate compact conformations of PGK induced by macromolecular crowding [12]. Now, by studying the competition of temperature, pressure and crowding on the energy landscape, we observe a costly barrier between two specific phases disappears, along a critical line on top of the isochore surface. As such, the current investigation demonstrations a richer ensemble of PGK states (Fig. 2) than our previous study [12].

To test our computational model, we observe structural transitions of PGK by fluorescence to construct the experimental $T$-$P$-$\phi$ phase diagram (Fig. 3). Experiment verifies the predicted existence of a critical point where $T_c$ moves to a lower temperature $T$ as the crowding volume fraction $\phi$ increases. Furthermore, we derive a critical line $T_c(\phi)$ using scaling arguments from polymer physics and present a unified phase diagram (Fig. 4) to investigate the underlying physical origin of such transition. As a consequence of being near the critical regime, PGK exhibits large structural fluctuations at physiologic conditions, which may be advantageous for enzymatic function. The current investigation is transforming the typical "structure-function" problem in proteins to a novel paradigm of a "structure-function-environment" relationship and is a step toward developing universal thermodynamic principles of protein folding in living cells.

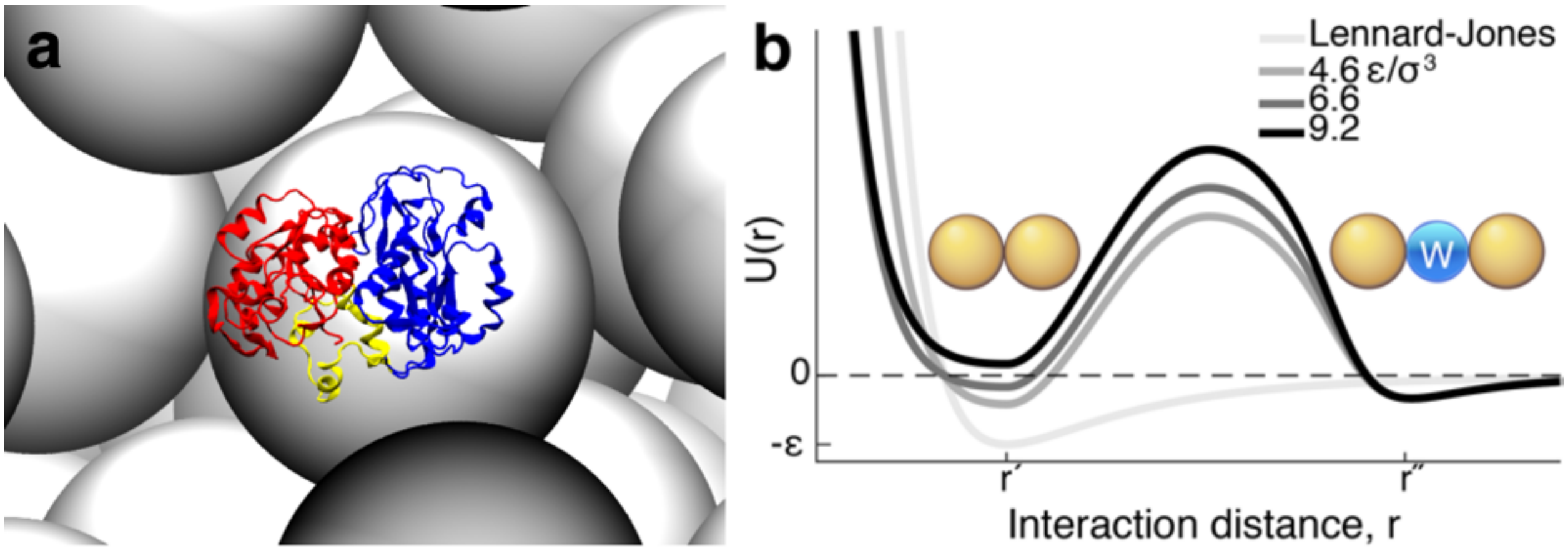

FIG 1. PGK surrounded by crowders and the desolvation potential between residues. (a) A snapshot from the coarse-grained molecular simulation of PGK's spherical compact state (Sph) surrounded by crowders (gray) at the volume fraction of 40%. N-, C-domain, and hinge are in red, blue, and yellow, respectively. (b) The pressure-dependent desolvation potential at $\sigma^3 P/\varepsilon$ = 4.6, 6.6, and 9.2, contains a *desolvation barrier* with a width ($|r' - r''|$) the size of a water molecule (blue). This incorporates the entropic cost of expelling a solvent molecule between two residues (gold). The Lennard-Jones potential is plotted in light grey for comparison.

## II. RESULTS AND DISCUSSION

### A. Computational $T$-$P$-$\phi$ phase diagram of PGK

We investigated the conformations of PGK, a large, 415 amino acid, two-domain protein ( [10] in section S1 for more information on PGK), in an environment containing Ficoll 70, which acts as a crowding agent mimicking cell-like excluded volume. Ficoll 70 is computationally modeled as a hard sphere, as it is known to be inert to proteins and behaves as a semirigid sphere {cite} [29,30]. From prior FRET (Förster Resonance Energy Transfer) experiments and molecular simulations, we gained knowledge of several PGK conformations that denote a phase diagram in the $\phi$-$T$ plane [12]. It includes four states: C (crystal structure), CC (collapsed crystal), Sph (spherically compact state), and U (unfolded structures). In the C state, there is a linker that separates the N-terminal and C-terminal domains, resembling an open "Pacman". The CC state is a closed "Pacman". The Sph state involves a twisting of one of the domains with respect to the other and becomes more spherical than the CC state. A complete description of the structures of these states is in Supplemental Material [10] section S3.

By changing hydrostatic pressure $P$ and volume fraction of crowders $\phi$ at several temperatures $T$, we have identified two new states on the $\phi$-$P$ isothermal phase plane (Fig. 2): I (folding intermediate), and SU (swollen compact unfolded structure). The criteria to define the six distinctive conformations are in Table S3.1 [10]. The I state is an ensemble of structures containing one folded domain (C-terminus) and one unfolded domain (N-terminus), making a specific prediction as to which domain is least stable on its own (N-terminal). SU is completely denatured but exhibits many water-mediated contacts [Fig. 1(b)]. Thus, the SU state is structurally more compact than the U state.

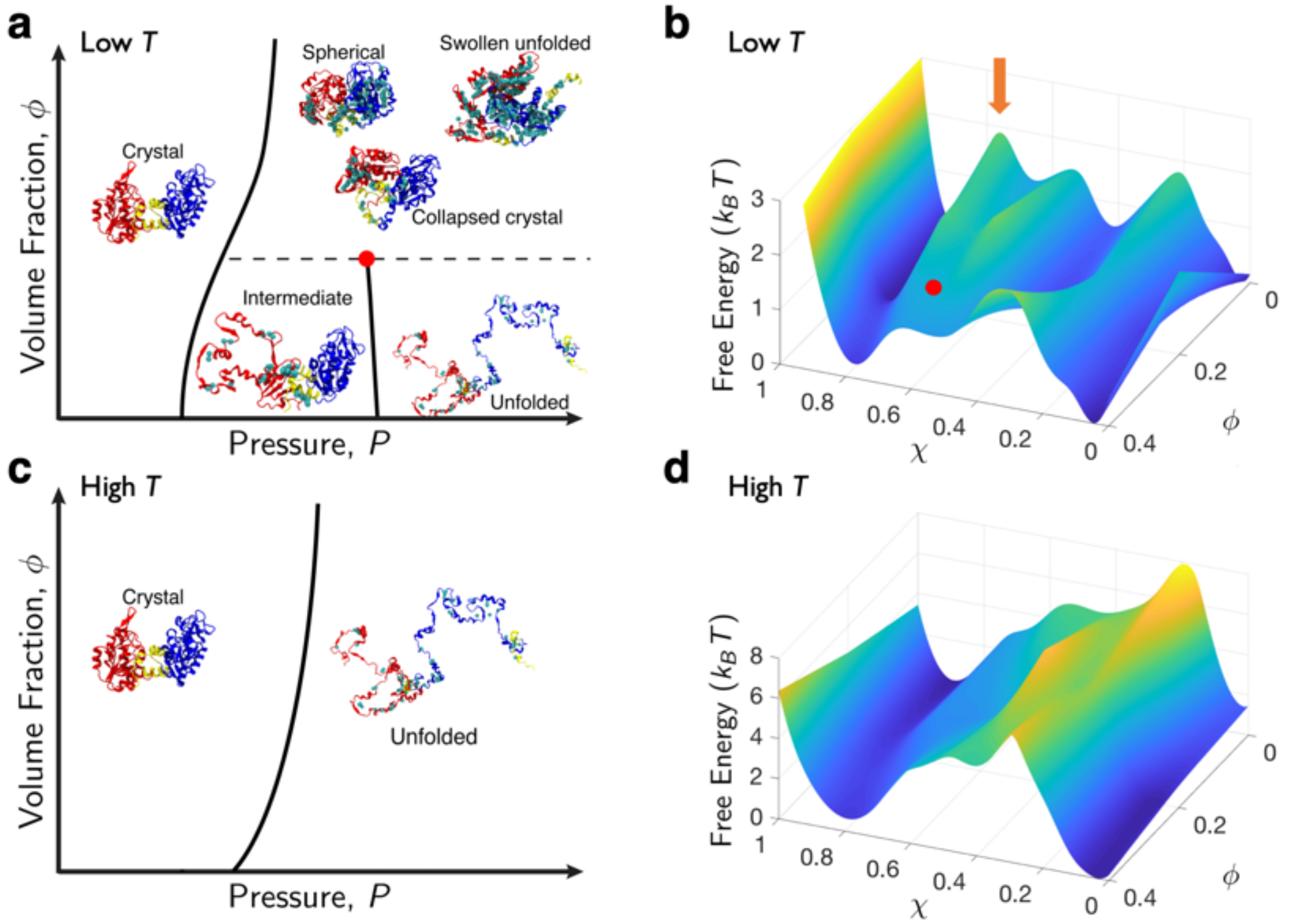

FIG 2. Solvation and crowding give rise to an intricate phase diagram of PGK. (a & c) Schematics of PGK's behavior in the crowding volume fraction-pressure ($\phi - P$) phase plane and (b & d) corresponding free energy with respect to the overlap $\chi$ and crowding volume fraction $\phi$ at the folding pressure at low (a & b) and high (c & d) temperatures. Solid lines represent the division between distinct configurational phases that are separated by a free energy barrier from simulations at $\phi = 0$, 0.2, and 0.4, and pressures from $\sigma^3 P/\varepsilon = 10^{-3}$ to 23. The dashed line (a) represents a continuous transition along $\phi$ and red dots (a & b) represent an approximate position of the critical points. The orange arrow (b) marks the peak of the barrier that diminishes until it disappears after the critical point. Collapsed crystal, spherical, and swollen unfolded states are indistinguishable in terms of free energy. These configurations were reconstructed from coarse-grained models to all-atomistic protein models for illustration purposes using SCAAL [31]. N-, C-domain, and hinge are in red, blue, and yellow, respectively. A cyan sphere was inserted in between residues to show water-mediated contacts.

The microscopic mechanism of the pressure-induced unfolding of PGK depends on the $T$ and $\phi$. Fig. 2 shows the $P$-$\phi$ phase plane at low $T$ in Fig. 2(a) and high $T$ in Fig. 2(c). At sufficiently low $T$ and $\phi = 0$ (no crowders) the unfolding of PGK is a multi-state transition between crystal state C [Fig. 2(a) & 2(b)] and unfolded state U via an intermediate state I. We capture the folding process using the overlap parameter $\chi$ ( [10] for definition), which characterizes similarity to the crystal structure, C state. $\chi$ ranges from 0 to 1 where 0 represents the C state. In Fig. 2(b), the I state has $\chi = 0.35$, and the U state has $\chi = 0.9$. The state I is a consequence of the heterogeneous distribution of cavities, causing uneven pressure-denaturation where N-domain unfolds, and C-domain remains intact. Since the total cavity volume of the N-terminal domain ($\approx$171 Å$^3$) is about a third larger than that of the C-terminal domain ($\approx$132 Å$^3$), the former is more vulnerable to high pressure. Moreover, two antiparallel β-strands *m* and *n* of the N-terminal domain are totally exposed to the solvent ( [10] in section S5 and Fig S1.1). Under high pressure, they act as a channel for water to fill the N-terminal domain's cavities.

At sufficiently high $\phi$ and low $T$ [Fig. 2(a), above the red critical point], there is only a single transition due to pressure between a crystal state and several compact states (Sph, CC, and SU) without the I state. The transition from C to Sph or CC states involves domain rearrangement when the linker "cracks" [11] and forms a disordered hinge. Whereas, high pressure competing with crowding gives rise to another compact unfolded conformations where up to half of the contacts becomes swollen with water that forms a "wet interface" (swollen unfolded states, SU). As the limited void formed by the density fluctuations of crowders inhibits extended conformations [32], the U state is unfavorable due to macromolecular crowding [31]. The protein only needs to subtly reduce its volume as it expels water molecules out of this wet core to return to the Sph or CC state from the SU state. There are effectively no barriers between the Sph, CC, and SU states, which are thus located in the same region of the phase diagram (see Fig. 2(a) and 2(b) at $\chi = 0.4$ to 0.8, and S5.2 [10]). This data supports the hypothesis that protein dynamics is governed by the solvent motion [33], and water inside the protein "lubricates" the transitions between conformations without significant free energy costs [25].

Similarly, at high $T$ [in Fig. 2(c) & 2(d)] ranging from low to high $\phi$, PGK also undergoes a single pressure-denaturation transition, but it is between the C and U states. Due to the increase in $T$, the U state is entropically more favorable than all other states. As such, the U state's entropy considerably compensates the C state's energy, causing an increase in the free energy barrier between $\chi = 0$ (C state) and 0.8-0.9 (U state) in Fig. 2(d).

Our model predicts from these $P$-$\phi$ slices at various $T$ that crowding makes the folding of PGK two-state, whereas lack of crowding produces a multi-state transition below a critical temperature $T_c$. Therefore, PGK undergoes a critical transition through by either of two directions on the $T$-$P$-$\phi$ phase space. One direction is by increasing $\phi$ at low $T$ and sufficiently high $P$ surpassing a critical volume fraction $\phi_c$ as shown in Fig. 2(a) at the red critical point. This transition is clearly seen by the diminishing of the free energy barrier in Fig. 2(b) pointed by an orange arrow, and $\phi_c$ is between 0.2 and 0.4. The second way is by increasing $T$ at low $\phi$ and sufficiently high $P$ surpassing a critical temperature $T_c$. Take $\phi = 0$ as an example; one of the free energy barriers in Fig. 2(b) pointed by an orange arrow diminishes. As a result, the multi-state free energy becomes two-state resembling the high $T$ free energy shown in Fig. 2(d). Thus, this suggests that the value of $T_c$ decreases as $\phi$ increases.

### B. Experimental *T-P-$\phi$* phase diagram of PGK

To validate the computed phase diagram, we measured the $P$-$T$ phase diagram of PGK at various Ficoll 70 crowder concentrations to obtain the full $P$-$T$-$\phi$ information experimentally (Fig. 3). While one cannot expect the exact temperatures and pressures to agree, identical topologies of the experimental phase diagrams validate the general conclusions from simulations. Changes of the states of PGK were detected by tryptophan fluorescence because tryptophan mean fluorescence wavelength is sensitive to water exposure as the protein unfolds. We scanned $T$ from 283 to 318 K at constant $P$, and $P$ from 0 to 250 MPa at constant $T$ with 0, 25, 50, 100, 150, and 200 mg/mL of Ficoll 70 concentration ($\phi = 0$ to $\approx 0.56$) to cover the complete phase diagram. Each transition produces a sigmoidal step in the plot of mean tryptophan fluorescence wavelength $\lambda_m$ vs. $P$ (Fig. S2.1 [10]).

In the absence of crowder at sufficiently low $T$ [Fig. 3(a), blue trace], there are two steps in $\lambda_m$ as a function of $P$, signaling two separate transitions among three states. These steps are straightforwardly revealed by plotting $\partial\lambda_m/\partial P$ and identifying peaks (see Fig 3(a), and Fig. S2.1 [10]). We assign the first peak to the C to I transition, the second to the I to U transition. At sufficiently high $T$, at $\geq 303$ K and $\approx$170 MPa, one of the peaks disappears [Fig. 3(c), blue trace], leaving only one transition between two states. We assign this to a direct transition from C to U, as shown in Fig. 2(c), corresponding to a critical point at $T_c = 306\pm3$ K. Finally, when crowder is added, $T_c$ moves to lower temperature, until the apparent three-state transition is no longer observed at all at 200 mg/ml Ficoll 70 [Fig. 3(a) & 3(c), red traces]. We assign this to the transition between C and SU/Sph/CC as shown in Fig. 2(a). Accurate transition midpoints ($T_m$, $P_m$, $\phi_m$) were obtained from each trace by fitting to sigmoidal two- or three-state models (solid curves in Fig. 3(a) & 3(c); see Appendix B; all data traces are shown [10] in section S4). Singular value decomposition analysis ( [10] in section S4) also strongly supports the conclusions obtained from analyzing $\lambda_m$.

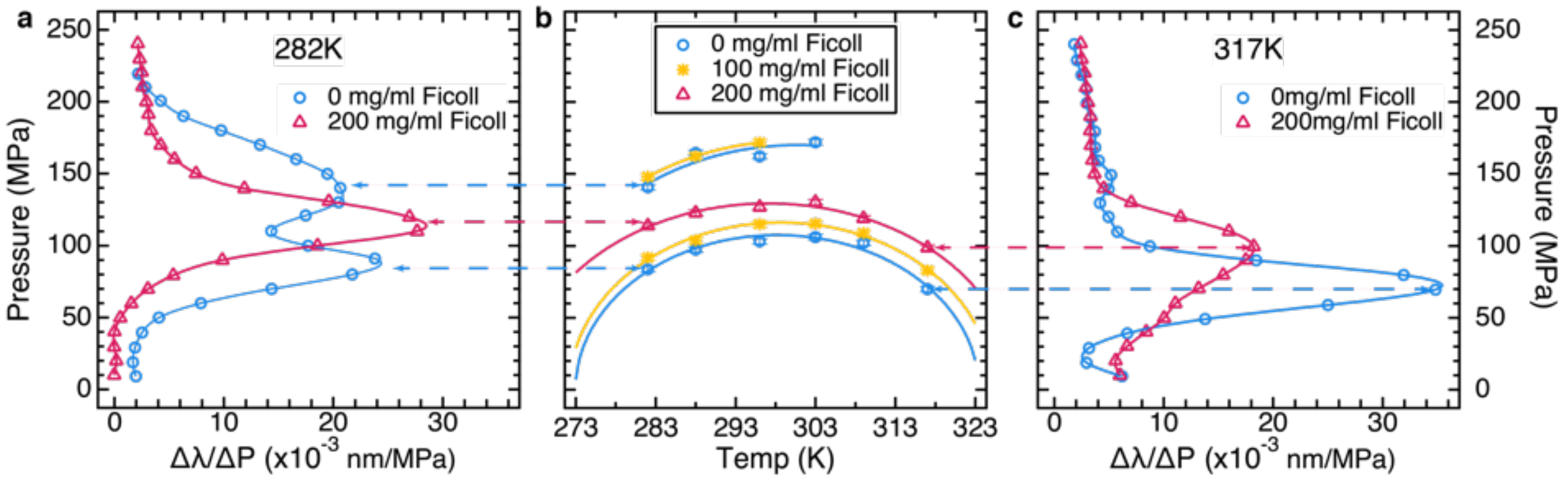

FIG 3. Experimental $P$-$T$-$\phi$ phase diagram of PGK (full data in [10]). (a) The derivative of the mean tryptophan fluorescence wavelength vs. pressure of PGK at 282 K calculated from fluorescence spectra. Two of six Ficoll 70 concentrations are shown. The markers show the data points and the solid line shows a cubic spline interpolation. The blue curve (0 mg/ml Ficoll 70) has two peaks as pressure increases, signifying two transitions; the magenta curve (200 mg/ml Ficoll 70) has only one peak point, signifying only one transition when pressure is applied. The dashed lines point from transition midpoints to the corresponding point in the phase diagram. (b) $P$-$T$ phase diagrams at several $\phi$ obtained by fitting the fluorescence data to obtain the inflection points of $\lambda_m(P)$ (peaks in the derivative $\partial\lambda_m/\partial P$). Three of six Ficoll 70 concentrations are shown. Circles represent midpoint pressures measured at 282, 288, 296, 303, 309 and 317K in absence of Ficoll 70 (0 mg/ml), asterisks represent transitions for the middle Ficoll 70 concentration (100 mg/ml) and triangles represent transitions for the highest Ficoll 70 concentration (200 mg/ml). At high $T$, or upon increasing Ficoll 70 concentration, the second (higher $P$) transition disappears, mapping out a critical point that moves to lower $T_c$ at higher Ficoll 70 concentration. Solid elliptical curves going through the circles are fits to Eq. (1) representing the $\Delta G = 0$ curves. (c) Equivalent data as in (a) at 317 K. Note that the second (higher $P$) transition is never present at high $T$.

We constructed $T$-$P$ planes of the phase diagram at all crowder concentrations as follows: First, the transition midpoints were plotted on $P$-$T$ slices at constant $\phi$, as shown in Fig. 3(b). These points correspond to zero free energy difference, $\Delta G = 0$, for the first-order transition, where concentrations of C and I, I and U, or C and U (depending on the location on the phase diagram) are equal. Then the transitions were fitted to Hawley's elliptical $P$-$T$ phase curve for proteins [34],

$$\Delta G(T,P) = \frac{1}{2}\Delta\beta(P-P_0)^2 + \Delta\alpha(T-T_0)(P-P_0) - \Delta C_P\left[T\left(\ln\frac{T}{T_0}-1\right)+T_0\right] + \Delta V_0(P-P_0) - \Delta S_0(T-T_0) + \Delta G_0, \quad (1)$$

at each value of $\phi$ (fits for all $\phi$ and parameter definitions in [10] section S4). Here $\Delta\beta$, $\Delta\alpha$, $\Delta C_P$, $\Delta V_0$, and $\Delta S_0$, signify changes in compressibility, thermal expansion coefficient, heat capacity, volume, and entropy, respectively. The resulting experimental phase diagram in Fig. 3(b) agrees with the computational data as both exhibit two pressure transitions at low $T$, one at high $T$, and a shift from two to three transitions at a value of $T_c$ that decreases with increased crowding.

The simulation predicts that in the state I, the N-terminus would be unfolded, and the C-terminus folded. We truncated the protein to the N-terminal domain and indeed found it to be unfolded with long tryptophan fluorescence wavelength and no cooperative transition (Fig. S4.4 [10]). It is known from the literature [35] that the C-terminal domain of PGK is stable by itself. These two observations combined strongly support the computational assignment of the I state with the N-terminal domain primarily unfolded, and the C-terminal domain mostly folded. Thus, experiment and simulations are in agreement both on the disappearance of the difference between two phases at high $T$ or high $\phi$, as well as the general structural features of the I state formed at low crowding.

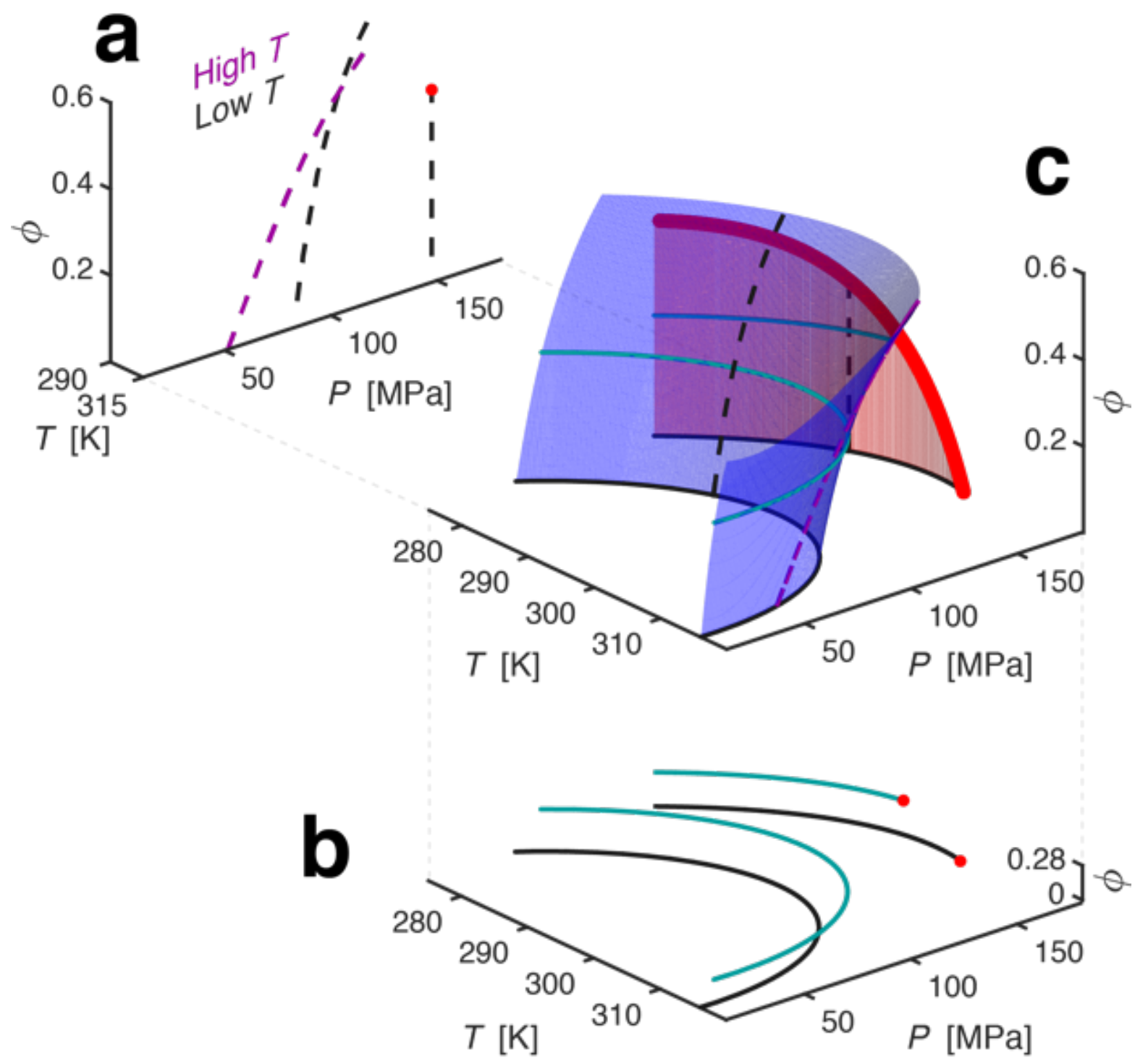

FIG 4. $T$-$P$-$\phi$ phase diagram of PGK from theory mapped onto the experimental data. (a) PGK's $\phi$-$P$ phase plane at high (magenta) and low (black) $T$. Dotted lines represent the division between distinct configurational phases. The red dot signifies a critical point. (b) Slices of $P$-$T$ phase diagram observed experimentally at no Ficoll 70 (black) and 100 mg/mL Ficoll 70 (cyan). Note that I-U coexistence curve terminates at the critical point (in red dots) and shifts $T_c$ to a lower temperature in the presence of Ficoll 70. Solid elliptical curves going through the circles are the fits representing the ΔG = 0 curves. (c) A $T$-$P$-$\phi$ phase diagram of PGK. The blue and red surfaces are the C-I (or C-U, depending on $T$ and $\phi$) and I-U coexistence surfaces, respectively. The dashed magenta and black line are the $\phi$-$P$ cross-section from Fig 4a, and the solid black and cyan are the $P$-$T$ cross-section from Fig 4b. The bold, red line bordering the red surface is the critical line.

### C. Unified *T-P-$\phi$* phase diagram of PGK

The three-dimensional (3D) *T-P-$\phi$* phase diagram in Fig. 4(c) presents a unified picture of the computational and experimental results. This unified phase diagram includes two surfaces: the blue surface represents C-I (or C-U, depending on $T$ and $\phi$) and the red surface represents I-U coexistence surfaces, respectively (the calculations of the surfaces can be found in Appendix C and [10] section S6). The projection of this 3D coexistence surface onto a 2D $\phi$-$P$ plane in Fig. 4(a) shows a low and high T slice as found computationally in Fig. 2. When projected on $P$-$T$ plane in Fig. 4(b), it shows a low $\phi$ and high $\phi$ slice as found experimentally in Fig. 3. As temperature increases, the second transition surface terminates at a critical line [bold red line on the red I-U coexistence surface in Fig. 4(c)]. As the crowding volume fraction increases, the critical point on each $P$-$T$ slice shifts towards lower temperatures. Thus, from the experiment, above $\phi = \phi_c \approx 0.5$ or $T = T_c^0 \approx 306$ K the pressure-induced folding transition contains only two apparent phases. Whereas at low $\phi$ and $T$, PGK exhibits apparent three-state folding.

To quantitate the unified phase diagram, we modified Hawley's theory by incorporating the free energy change due to crowding, as is similarly treated in Minton's theory [16], to construct the first transition surface [blue surface in Fig. 4(c)]. As for the critical line on the second transition surface [in red in Fig 4(c)], we used scaling arguments to derive the equation for the critical line,

$$P - P_c = a_1\big(T - T_c(\phi)\big) + a_2\big(T - T_c(\phi)\big)^2 + \mathcal{O}(\Delta T^3), \qquad (2)$$

by treating the protein's U to I transition similar to in the coil-globule transition of theory [36,37]. Here,

$$T_c(\phi) = T_c^0\left(1 - \frac{\phi}{\phi_c}\right)^{\gamma}, \qquad (3)$$

where $T_c^0$ is the critical temperature without crowding, $\phi_c$ is the critical crowding volume fraction, $P_c$ ($\approx 170$ MPa) is the critical pressure taken from our experiment at $T_c^0$, and $a_1 = \left.\frac{dP}{dT}\right|_{T=T_c^0}$ and $a_2 = \left.\frac{d^2P}{dT^2}\right|_{T=T_c^0}$. From the fitting to experimental critical points at all slices of $\phi$, we found $\gamma = 0.40 \pm 0.01$, which is the predicted scaling exponent of a polymer collapse due to crowders, $\gamma = 2/5$ [38,39] (see Appendix C and [10] section S6). From this phase diagram, we can see the protein moves through a diverse phase space, suggesting different folding mechanisms that depend on how the phase diagram is traced out [40,41].

### D. Consequences of criticality

In Fig. 5, we explore the impact of $P$ and $\phi$ on the folding of PGK. The consequences of the critical regime are revealed by the ensemble distributions of the cavity volume (conjugate variable of $P$) and co-volume (conjugate variable of $\phi$) [15,16] from our simulations (also see Fig. S5.1 [10]). In the critical regime, small perturbations in crowding $\phi$, $P$, or $T$ will significantly affect the system.

We investigate the response of the ensembles near the conformational distribution of structures close to the critical region by comparing the cavity volume fluctuations ($\delta V^2 = \langle V^2\rangle - \langle V\rangle^2$) (or proportionally, the compressibility) and structural fluctuations ($\delta\chi^2 = \langle\chi^2\rangle - \langle\chi\rangle^2$) in the presence and absence of crowding agent. PGK has larger $\delta V$ [Fig. 5(a)] at $\phi = 0.4$ with a peak at 6.6 $\varepsilon/\sigma^3$ than that of $\phi = 0$. We suspected that the critical regime is between $\phi = 0.2$ and 0.4 and between pressures 4.6 $\varepsilon/\sigma^3$ and 6.6 $\varepsilon/\sigma^3$ at a temperature of 0.97 $\varepsilon/k_BT$ in the computational model, which qualitatively agrees with the experiment. Even though $\delta\chi$ is large in the presence of crowders, structures lie in a narrow range of co-volumes, making them indistinguishable to macromolecular crowding effects if shape can be neglected to the 0th order [Fig. 5(b)]. A sample of the diverse structures with similar cavity volumes and co-volumes are shown in Fig. 5(c).

Not only does crowding shift the population of structures to more compact states such as CC or Sph (Fig. 2 and [12]), where the two ligand binding sites (for ADP and 1,3-DPG) come into close proximity of each other, but it also increases the structural fluctuations of the compact states by bringing PGK closer to the critical regime, as shown in Fig. 5. Both of these properties would most likely facilitate enzymatic activity. This is corroborated by previous FRET experiments that show an increase in PGK's enzymatic activity as Ficoll 70 concentration increases [12]. These results suggest that criticality assists the enzymatic function of a protein.

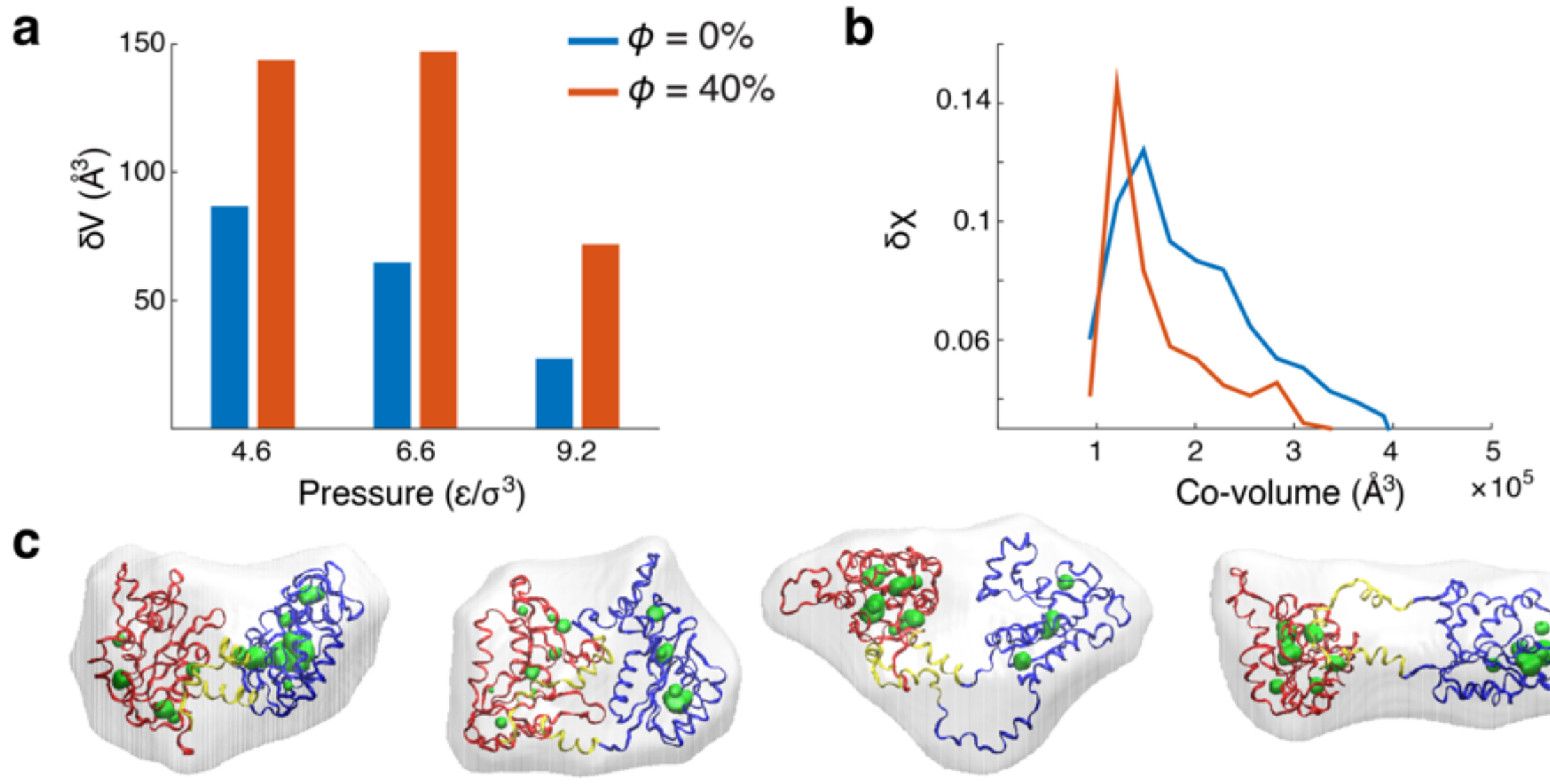

FIG 5. Cavity volume and structural fluctuations near critical regime. (a) Cavity volume fluctuations, $\delta V^2 = \langle V^2 \rangle - \langle V \rangle^2$, (or proportionally compressibility) of PGK at $k_B T/\varepsilon = 0.97$ and $\sigma^3 P/\varepsilon = 4.6$, 6.6, and 9.2 with (orange) and without (blue) crowding. (b) Overlap fluctuations, $\delta\chi^2 = \langle \chi^2 \rangle - \langle \chi \rangle^2$, as a function of co-volume at pressure $\sigma^3 P/\varepsilon = 6.6$. (c) Conformations from the ensemble in presence of crowding at $k_B T/\varepsilon = 0.97$ and $\sigma^3 P/\varepsilon = 6.6$ with co-volumes $\approx 1.1 \times 10^5$ Å$^3$ and cavities $\approx 200$ Å$^3$. Most left conformation is a crystal state. The following structures from left to right have a $\chi = 0.31$, 0.38 and 0.48. N-, C-domain, and hinge are in red, blue, and yellow, respectively. Co-volumes are shown as translucent surfaces surrounding the protein and cavity surfaces are shown in green.

## III. CONCLUSION

In summary, we have shown direct evidence of equilibrium critical-like behavior on the *T*-*P*-$\phi$ phase diagram of a protein by computational simulations, by fluorescence spectroscopy, and by a theoretical argument based on polymer physics. Despite the simplicity of the computational and theoretical model, all three different approaches agree with one another, validating the trends on the *T*-*P*-$\phi$ phase diagram and presence of the critical regime. Above the critical line in Fig. 4(c) (by increasing *T*, $\phi$, or both at $P_c \approx 170$ MPa), the difference between the I and U phases disappears. This is due to the loss of the free energy barrier between the two phases [orange arrow in Fig 2(b)] and is reaffirmed by the high-pressure fluorescence measurements (Fig 3).

What is the origin of the critical behavior in proteins? To answer this question, two concepts need to be rationalized together. Firstly, proteins are biopolymers that often undergo an abrupt or first-order-like transition to a compacted folded state from an expanded unfolded state or coil at a folding temperature, $T_F$. Secondly, the coil-globule transition seen in other polymers is a continuous transition at a specific temperature called $\theta$-temperature, $T_\theta$ [36,37]. Therefore, the first order transition in protein folding must be occurring near the collapse transition ($T_F \approx T_\theta$), meaning it normally is *tricritical* [7]. In the current system, the pressure perturbation may cause $T_F \neq T_\theta$, separating the continuous and first-order transitions. When going from a continuous to a first-order transition, there are signatures of passing through a critical point [42,43]. Finally, when $\phi$ is high ($\phi > \phi_c$), the protein is already collapsed even when it is unfolded. Our theoretical model in Eq. (3) and Appendix C (also [10] section S6) captures this postulation of the basis of criticality in proteins.

Furthermore, our computational and experimental results are in accord with the capillarity picture of folding [44], which posits a wetting interface between folded and unfolded parts of a protein, giving rise to a diverse phase space. Strong macromolecular crowding, which drives conformational changes to favor compact states, roughens that wetting interface, allowing cavities to spread throughout the conformation of the protein, with two major consequences. A roughened interface reduces activation barriers for folding, driving multi-state transitions towards apparent two-state transitions. It also creates a critical state where heterogeneous conformations coexist, as the front of wetting interface moves across the protein.

We conclude that large structural fluctuations (Fig 5) and merging of protein phases are consequences of being close to a critical point [Fig 4(c)]. At such a point, the barrier separating states vanishes (here: between I and U). Critical behavior has been proposed for protein folding at the onset of downhill folding [8], but its manifestation has been challenging to demonstrate computationally and experimentally [45]. Macromolecular crowding shifts the critical point to a lower temperature [Eq. (3)], indicating that such criticality could be physiologically important [3,4]: a protein near a critical regime could access a wide range of conformations without significant activation barriers for functional purposes inside the cell.

Further work will be needed to provide stronger evidence for the universality of critical behavior in proteins. Due to their complexity, proteins are not like other conventional condensed matter systems, and conventional tools, such as finite size scaling [46] or renormalization group theory [47], are not clearly applicable. The current investigation is a starting point toward developing universal principles of protein folding relevant to the environmental perturbations inside living cells and is an inspiration to create new tools to understand critical phenomena in these complex systems.

## ACKNOWLEDGMENTS

We thank Prof. Dave Thirumalai, Prof. Vassiliy Lubchenko, Prof. Greg Morrison, Prof. Peter G. Wolynes, Anna Jean Wirth, and Greg P. Gasic for helpful and stimulating discussions. M.B.P. was a Howard Hughes Medical Institute International Student Research Fellow, and then a Helen Hay Whitney Foundation Postdoctoral Fellow. A.G.G is supported by a training fellowship on the Houston Area Molecular Biophysics Program (T32 GM008280). M.S.C. and A.G.G. were funded by the National Science Foundation (MCB-1412532, PHY-1427654, ACI-1531814), and thank computing resources from the Center for Advanced Computing and Data Systems at UH. M.G. and M.B.P. were funded by NIH grant GM093318.

## APPENDIX A: SIMULATION MODEL

Our simulations use a structure-based model, which is minimalist protein model ("beads on a chain") that incorporates experimentally derived structural information [48], to investigate the mechanism of protein folding dynamics optimally. The emergence of structure and function from a protein sequence makes the modeling of proteins from first principle (*ab initio* models) computationally and theoretically prohibitive. Therefore, experimentally derived structural information is needed (even in models termed "all-atom", which refine *ab initio* force field parameters to fit experimentally known structures) to capture key features in protein folding and dynamics [49]. A structure-based model is often utilized as the "ideal gas" of protein folding for the investigation of a wide range of folding mechanisms [2,50]. This model renders an energy landscape [51] with minimal frustration and contains a dominant basin of attraction, corresponding to an experimentally determined configuration [52]. As such, the model carries the bonus of being computationally inexpensive, enabling long-timescale simulations to be obtained for a large protein and macromolecular crowding system. Long-timescale simulations are also crucial for high-pressure unfolding since pressure unfolds proteins at an order of magnitude (or more) slower than heat unfolding; therefore, structure-based, minimalist-model simulations provide statistically significant results. Lastly, structure-based models tend to capture unfolded protein scaling laws better than all-atom models [53], which is necessary to characterize the various non-crystal states of PGK correctly.

Similar to adding specific complexity to the ideal gas model to study specific phenomena, we add the desolvation barrier [25] to the native interactions that accounts for the free energy cost to expel a water molecule in the first hydration shell between two hydrophobic residues [54] to study pressure unfolding, leading to the appearance of a partially folded intermediate. The use of this model has been validated in other systems [28]. The Hamiltonian of this structure-based protein model is as follows:

$$\begin{aligned}\mathcal{H}_p(\Gamma,\Gamma^0) = &\sum_{i<j} K_r\,(r_{ij}-r_{ij}^0)^2\delta_{j,i+1} + \sum_{i\in\text{angles}} K_\theta\,(\theta_i-\theta_i^0)^2 \\ &+ \sum_{i\in\text{dihedrals}} K_\phi\left(\{1-\cos[\phi_i-\phi_i^0]\}+\frac{1}{2}\{1-\cos[3(\phi_i-\phi_i^0)]\}\right) \\ &+ \sum_{\substack{i,j\in\text{native}\\|i-j|>4}} U(r_{ij}) + \sum_{i,j\notin\text{native}} \epsilon\left(\frac{\sigma}{r_{ij}}\right)^{12} \end{aligned} \tag{4}$$

where $\Gamma$ is a configuration of the set $r, \theta, \phi$. The $r_{ij}$ term is the distance between $i^{\text{th}}$ and $j^{\text{th}}$ residues, $\theta$ is the angle between three consecutive beads, and $\phi$ is the dihedral angle defined over four sequential residues. $\delta$ is the Kronecker delta function. $\Gamma^0=\{\{r^0\},\{\theta^0\},\{\phi^0\}\}$ is obtained from the crystal structure configuration. Lastly, $U(r_{ij})$ is desolvation potential. Crowders are modeled as hard spheres. The complete descriptions of a structure-based protein model, desolvation potential, and simulations of PGK in a periodic cubic box of Ficoll 70 are provided in the Supplemental Material [10]. All simulations were performed using GROMACS 2016.3 molecular dynamics software [55].

## APPENDIX B: HIGH PRESSURE FLUORESCENCE EXPERIMENT METHODS

Fluorescence experiments were carried out at high pressure using a high-pressure cell (ISS High-Pressure Cell System) on a fluorimeter (JASCO FP8300). A computerized high-pressure generator (Pressure BioSciences Inc. HUB440) was used to pressurize the fluorescence cell. We used a rectangular quartz cuvette with a path length of 6 mm, and deionized water was used as the pressurizing fluid. The pressure was raised from 0.1 MPa to 250 MPa at a rate of 10 MPa/min and held at intervals of 10 MPa for a 5 min wait time to allow sample equilibration. Fluorescence spectra from 300 to 450 nm were acquired in the middle of the wait time and an in-built proportional-integral-derivative (PID) feedback loop was used to obtain accurate pressures (within 5 bars of target pressure). The temperature was controlled using an external water-circulating bath. To construct a complete $P$-$T$-$\phi$ phase diagram, fluorescence measurements were done at 6 different Ficoll 70 concentrations ([Ficoll 70] = 0, 25, 50, 100, 150 and 200 mg/ml), each at 6 different temperatures $T$ ranging from 282K to 317K (9°C to 44°C). Equilibrium traces of mean fluorescence wavelength vs. pressure [e.g., Fig. 3(a)] were fit to a two-state or three-state thermodynamic model (see Supplementary Material [10]), depending on whether the derivative (Fig. S1.2 [10]) of the titration curve identified one or two transitions (An inflection point in the fluorescence vs. $P$ trace at given $T$ and [Ficoll 70] produces a peak in the derivative). The fitted transition midpoints ($P$, $T$, $\phi$) were then plotted in a phase diagram (e.g., Fig. 3(b)) and fitted to Eq. (1). See [10] for complete data and fitting parameters.

## APPENDIX C: CONSTRUCTION OF PHASE DIAGRAM

We derived the critical line [Eq. (2) & (3); red line in Fig 4(c)] on the $T$-$P$-$\phi$ phase diagram using arguments based on the coil-globule transition [36,37] of a polymer. Beginning with a Landau-Ginsberg free energy [56], $F = -r(T,\phi)\Psi^2 + u\Psi^4 + F_0$, to describe the critical transition, where $\Psi$ is the order parameter, which is a scaled and shifted $R_g$ (radius of gyration) so that $\Psi = -\Psi_0$ for the I state and $\Psi = +\Psi_0$ for the U state. Since pressure is only involved with the first-order transitions, it can be ignored for now. At the critical temperature, the barrier between the I and U states vanishes, meaning $r = 0$; therefore, a reasonable function is $r(T,\phi) = -r_0[T - T_c(\phi)]$, where the critical temperature $T_c$ is a function of $\phi$, and $r_0$ is positive constant. To find the $\phi$ dependence of $T_c$, we used the scaling relationship

$$\frac{R_g(\phi)^2}{R_g(0)^2} \sim (1 - c_0\phi)^\gamma, \tag{5}$$

which relates $R_g$ at a given $\phi$ to $R_g$ without crowders for the collapse of a coil to globule transition [38]. The scaling exponent $\gamma$, is shown to be 2/5 in Refs [38,39]. Since the collapse of the polymer, or in the current case the protein, is dependent on $\phi$, and since $\Psi^2 \sim R_g(\phi)^2/R_g(0)^2$, the critical temperature $T_c(\phi)$ causing the free energy barrier between I and U to disappear must also scale as Eq. (5), giving Eq. (3) (see [10] section S6 for more details). We fit Eq. (3) to the experimental critical point values at all Ficoll 70 concentrations to find $\gamma$ and $\phi_c$. We fit Eq. (2) to experimental values of the I to U transition surface to find the Taylor expansion coefficients.

Lastly, we modified Hawley's equation [34] to fit the C to I (or U, depending on $T$ and $\phi$) transition surface (in blue in Fig 4(c)) by adding a $\phi$-dependent $\Delta G_{crowd}(\phi)$ term to Eq. (1),

$$\Delta G_{crowd}(\phi) = g\left(\frac{\phi}{1-\phi}\right) + \mathcal{O}(\phi^2), \tag{6}$$

making the 3D free energy change $\Delta G(T, P, \phi) = \Delta G(T, P) + \Delta G_{crowd}(\phi)$. This term is similar to Minton's theory [16], which treats the folded and unfolded proteins as effective hard spheres and employs scaled particle theory (SPT) to estimate the change in folding free energy as the difference between the insertion free energy for the folded and the unfolded states. Eq. (6) adds one more fitting parameter, *g*, to the total free energy change compared to Eq. (1).

## Supplemental Material

# Critical phenomena in the temperature-pressure-crowding phase diagram of a protein

Andrei G. Gasic,[1,2,*] Mayank M. Boob,[3,*] Maxim B. Prigozhin,[4,†] Dirar Homouz,[1,2,5]
Caleb M. Daugherty,[1,2] Martin Gruebele,[3,4,6,‡] and Margaret S. Cheung[1,2,§]

[1] *University of Houston, Department of Physics, Houston, Texas, 77204, United States.*

[2] *Center for Theoretical Biological Physics, Rice University, 77005, United States.*

[3] *Center for Biophysics and Quantitative Biology, University of Illinois at Urbana-Champaign, Champaign, IL 61801, United States.*

[4] *Department of Chemistry, University of Illinois at Urbana-Champaign, Champaign, IL, 61801, United States.*

[5] *Khalifa University of Science and Technology, Department of Physics, P.O. Box 127788, Abu Dhabi, United Arab Emirates.*

[6] *Department of Physics and Beckman Institute for Advanced Science and Technology, University of Illinois at Urbana-Champaign, Champaign, IL 61801, United States.*

[*] These authors contributed equally to this work.
Corresponding author contact information:
[‡] mgruebel@illinois.edu
[§] mscheung@uh.edu

## Table of Contents

## S1. Structure and Function of yeast phosphoglycerate kinase (PGK)

Yeast phosphoglycerate kinase [1] (PGK) is a 415-residue protein with two domains in which the N-terminal domain (residues 1–185) is slightly smaller than the C-terminal domains (residues 200–389). The domains are connected by a small α-helix (residues 186–199) together with the last 24 residues of the C-terminal tail (residues 390–415), which comprises the linker or "hinge" between the two domains [2]. PGK reversibly catalyzes a reaction step in the glycolysis pathway, in which a phosphate group is transferred from 1,3-bisphosphoglycerate (1,3-BPG) to adenosine diphosphate (ADP) to produce adenosine triphosphate (ATP). 1,3-BPG and ADP bind to the N- and C-terminal domains, respectively. Our work in 2010 shows that rather than a hinge-bending motion [3], the population shifts favoring compact states of PGK in which the two domains are in proximity to bring 1,3-BPG and ADP close to each other facilitates the transfer of the phosphate group in a cellular environment [4]. The molecular mass is about 45 kD. Table S1.1 shows residues of PGK's secondary structures and Fig S1.1 is a corresponding 2-dimentional schematic.

**Table S1.1** Constituent Residues of Secondary Elements in PGK (adapted from ref [5]). Parallel β-strands are labeled sequentially by uppercase letters. Antiparallel β-strands are labeled with lowercase letters. Helices are numbered in sequential order. N-domain: residues 1-185, C-domain: residues 200-389, Linker: residues 186-199 and 390-415, and Binding site: residues 23, 25, 38, 62, 121, 164, 211, 235, 236, 239, 310, 334-336, 338-341, 370-373.

| N-domain | | | | C-domain | | | | Linker | |
|---|---|---|---|---|---|---|---|---|---|
| β-strands | | α-helices | | β-strands | | α-helices | | α-helices | |
| A | 16-23 | 1 | 36-52 | G | 205-210 | 8 | 217-227 | 7 | 185-199 |
| B | 56-63 | 2 | 76-88 | H | 229-234 | 9 | 236-246 | 14 | 394-401 |
| C | 89-94 | 3 | 100-108 | I | 275-280 | 10 | 257-273 | 15 | 406-410 |
| D | 113-118 | 4 | 141-154 | o | 282-286 | 11 | 315-327 | | |
| m | 129-132 | 5 | 163-167 | p | 294-299 | 12 | 346-362 | | |
| n | 135-139 | 6 | 171-174 | J | 330-334 | 13 | 371-380 | | |
| E | 157-162 | | | K | 365-369 | | | | |
| F | 180-184 | | | L | 388-390 | | | | |

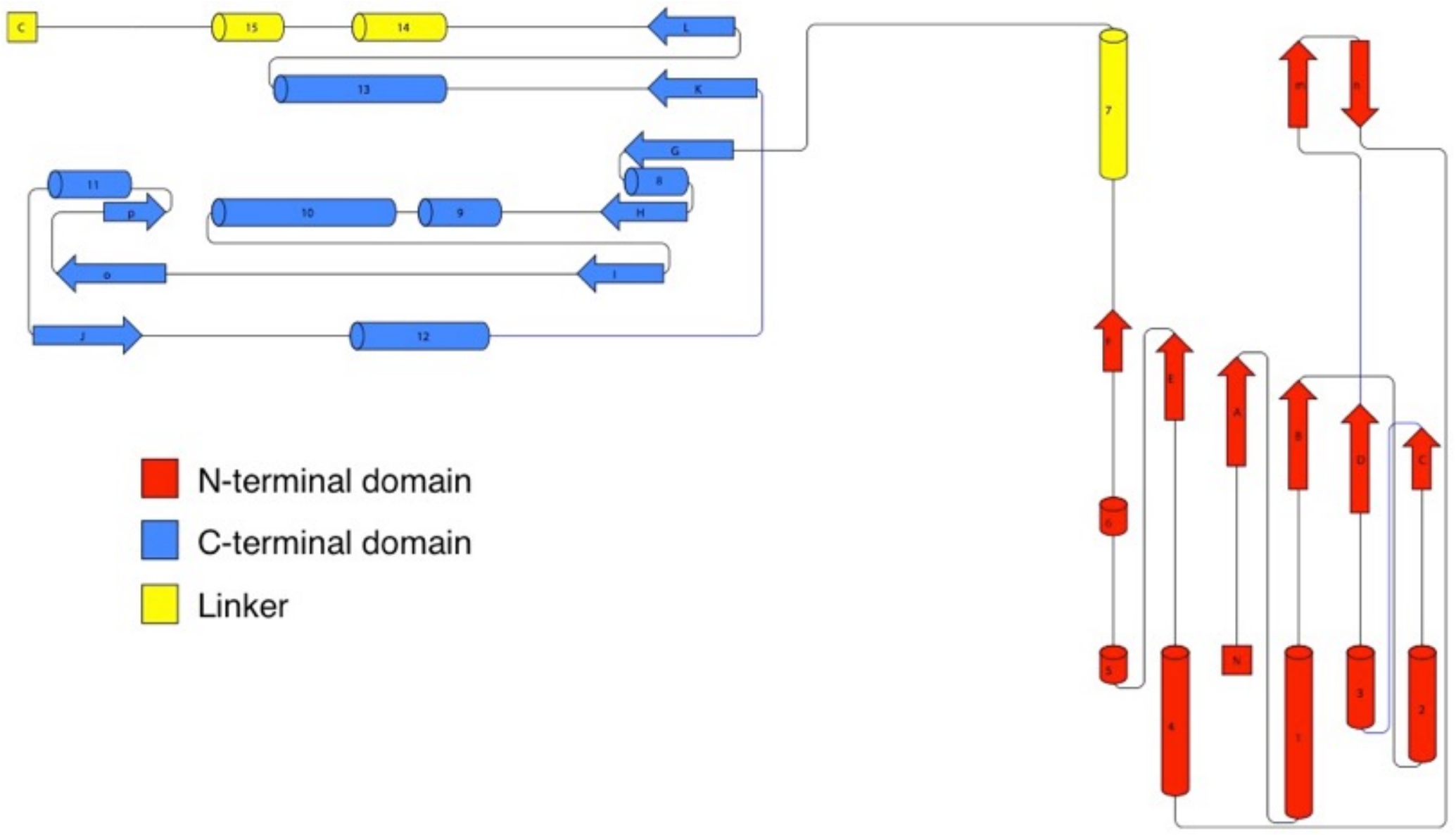

**Figure S1.1** 2D schematic representation of PGK. Labels corresponded to secondary structures in Table S1.1

## S2. Supplementary methods

### S2.1 Experimental methods

**Protein expression** Yeast phosphoglycerate kinase (PGK) was expressed as described previously [6]. Briefly, the PGK gene in pET28b vector was transformed into BL21-CodonPlus(DE3)-RIPL *E. coli* cells (Agilent) and selected overnight on kanamycin agar plates. A single colony was used to inoculate up to 10 L of LB culture media, which was grown at 37 °C until O. D. reached 0.6-1.0. The culture was then induced with isopropyl β-D-thiogalactopyranoside (IPTG) of 1 mM final concentration for approximately 8 hours at 23 °C. Cells were spun down for 10 min at 5000 rpm. The pellets were resuspended in a lysis buffer consisting of 50 mM $Na_2HPO_4$, 500 mM NaCl, and 10 mM imidazole, pH = 8.0. The cells were homogenized by sonication and the resulting solution was cleared by syringe filtration with 0.45 μm filters. The solution was then applied onto a Ni-NTA His-bind column (Novagen Inc.). The column was eluted by a gradient of imidazole buffers from 10 mM to 300 mM. Protein fractions were confirmed by gel electrophoresis and dialysed against 20 mM sodium phosphate buffer, pH = 7.0. Unless otherwise noted, measurements were done with 66 μM PGK in 20 mM sodium phosphate at pH 7.0, 10 mM dithiothreitol (DTT), and 10 mM ethylenediaminetetraacetic acid (EDTA). All chemicals were obtained from Sigma-Aldrich and used without further purification.

**Pressure denaturation monitored by fluorescence spectroscopy** Fluorescence spectra were measured using a Cary Eclipse fluorescence spectrophotometer (Varian). Excitation and emission slit widths were 5 nm each, the excitation wavelength was 280 nm, and the scan rate was 120 nm/min. Sample concentration was 66 μM unless otherwise specified. The sample was pressurized with a high-pressure cell (ISS). We used a rectangular quartz cuvette with a path length of 4 mm. Spectrophotometric grade ethyl alcohol (95.0 %, A.C.S. reagent; Acros Organics) was used as pressurization fluid. Temperature denaturation monitored by fluorescence spectroscopy was done in a similar manner but using a temperature controller instead of a pressurization cell. For a temperature melt at 50 MPa and for pressure melts at various temperatures, a water-circulating bath was used to control temperature and pressure simultaneously. The mean wavelength shown in figures is the weighted average of the fluorescence spectra [7]:

$$\langle \lambda \rangle = \frac{\int d\lambda\, \lambda\, I(\lambda)}{\int d\lambda I(\lambda)}$$

Equilibrium traces were fit using a thermodynamic two-state and three-state model. The total signal was represented as a linear combination of Boltzmann distributions of each state, where each state was assumed to have a linear baseline. For example, for a two-state case, signal, *S*, as a function of pressure, *P*, is:

$$S(P) = \frac{a_i + a_s(P-P_m) + (b_i + b_s(P-P_m))e^{\Delta V(P-P_m)/RT}}{1+e^{\Delta V(P-P_m)/RT}},$$

where $a_i$, $a_s$, $b_i$, and $b_s$ are linear baselines of the two states, $\Delta V$ is the change in volume, $P_m$ is the transition pressure, $R$ is the gas constant, and $T$ is temperature. Similarly, a three-state case as a function of pressure, *P*, will be:

$$S(P) = \frac{a_i + a_s(P-P_{m1}) + (b_i + b_s(P-P_{m1}))e^{\frac{\Delta V_1(P-P_{m1})}{RT}} + (c_i + c_s(P-P_{m2}))e^{\Delta V_2(P-P_{m2})/RT}}{1 + e^{\Delta V_1(P-P_{m1})/RT} + e^{\Delta V_2(P-P_{m2})/RT}}$$

where, $a_i$, $a_s$, $b_i$, $b_s$, $c_i$ and $c_s$ are linear baselines of the three states, $\Delta V_1$ and $\Delta V_2$ are the change in volume, $P_{m1}$ and $P_{m2}$ are the two transitions pressures. Determining whether a two-state model or a three-state model was needed to fit the data was done as follows: $\partial\langle\lambda\rangle/\partial P$ was evaluated as shown in Figure S2.1 below. When the transition derivative showed two large peaks or a peak with a shoulder, a three-state model was fitted; if only a single large peak was evident in the transition derivative, a two-state model was fitted. The three-state transition moves to lower temperature as the Ficoll 70 concentration increases.

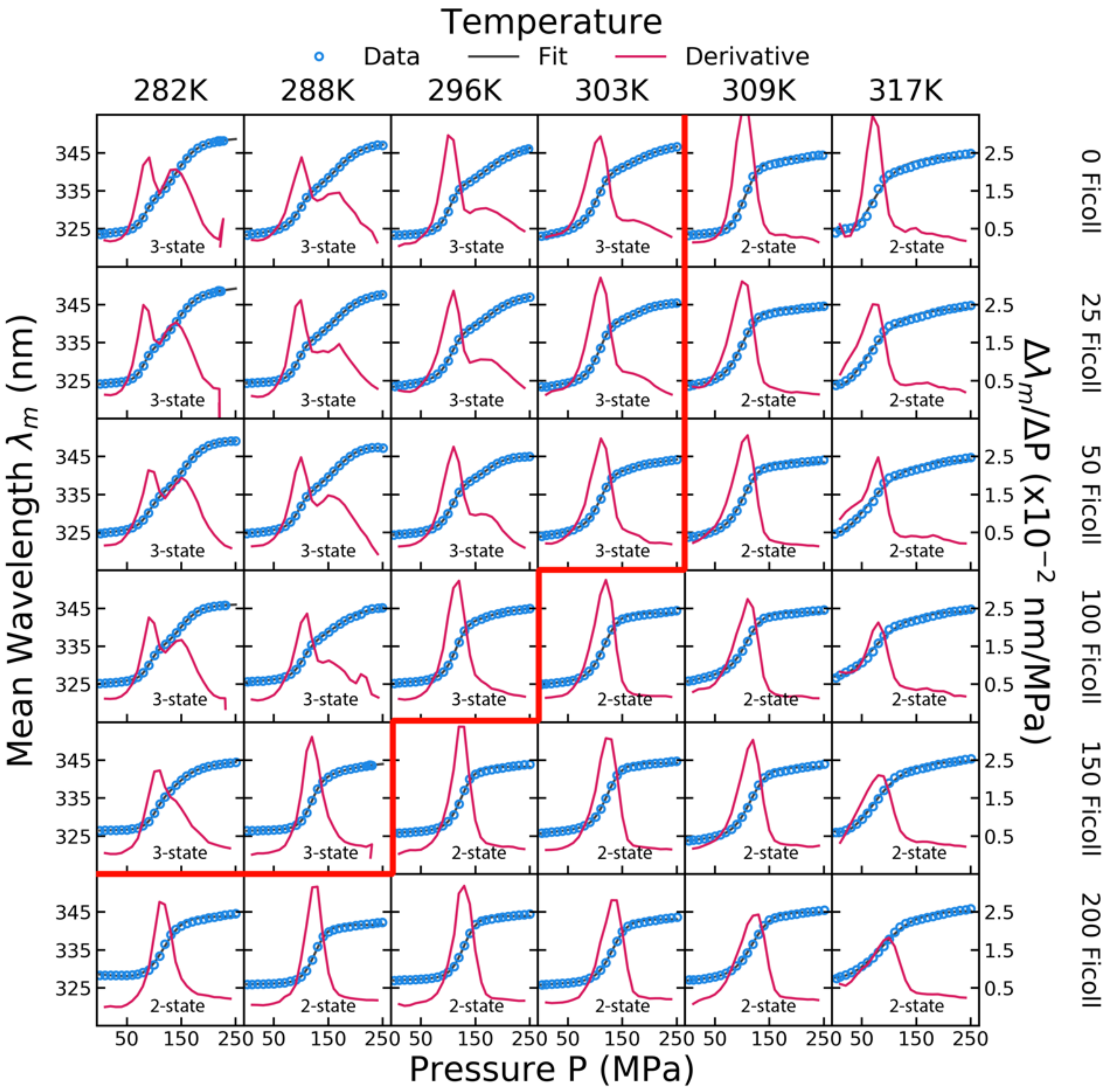

**Figure S2.1** Combined data set for all the pressure melts performed at different concentration of Ficoll 70 in units of mg/ml and different temperatures along with fit (plotted on the left Y axis of subplots) and derivative of the data (plotted on the right Y axis of each subplot). The derivative was used to distinguish between a two-state and three-state model. When there is a single peak in the derivative (signifying one transition), data were fitted to a two-state model. When there are two peaks in the derivative (signifying two transitions), data were fitted to a three-state model. The red line is shown as a guide to separate the data showing 3-state behavior vs the date showing 2-state behavior.

**Protein sequence:**

```
MGSSHHHHHH SSGLVPRGSH MSLSSKLSVQ DLDLKDKRVF IRVDFNVPLD
GKKITSNQRI VAALPTIKYV LEHHPRYVVL ASHLGRPNGE RNEKYSLAPV
AKELQSLLGK DVTFLNDCVG PEVEAAVKAS APGSVILLEN LRWHIEEEGS
RKVDGQKVKA SKEDVQKFRH ELSSLADVYI NDAFGTAHRA HSSMVGFDLP
QRAAGFLLEK ELKYFGKALE NPTRPFLAIL GGAKVADKIQ LIDNLLDKVD
SIIIGGGMAF TFKKVLENTE IGDSIFDKAG AEIVPKLMEK AKAKGVEVVL
PVDFIIADAF SADANTKTVT DKEGIPAGFQ GLDNGPESRK LFAATVAKAK
TIVFNGPPGV FEFEKFAAGT KALLDEVVKS SAAGNTVIIG GGDTATVAKK
YGVTDKISHV STGGGASLEL LEGKELPGVA FLSEKK
```

**S2.2 Theoretical model**

In this study, we use an off-lattice $C_\alpha$ minimalist structure-based model of yeast Phosphoglycerate kinase (PGK) from the Protein Data Bank (PDB ID: 1QPG [5]). The mutation R65Q of this structure was reverted to the original, and three additional mutations were made at positions Y122W/W308F/W333F to match the experimental preparation exactly. We employed energy minimization and simulated annealing to computationally remove steric clashes. We use contacts of structural units (CSU) [8] software to define native contacts. CSU software takes into account all the topological constraints of our structure. The model incorporates a structure-based interactions on residues in close proximity in the PDB crystal structure that are modified to incorporate a water-mediated interactions [9]. The semirigid, neutral, macromolecular crowding agent, Ficoll 70, is modeled as a hard-sphere [10]. The Hamiltonian ($\mathcal{H}_{tot}$) of the entire system consisting of PGK and spherical crowders is $\mathcal{H}_{tot} = \mathcal{H}_p + \mathcal{H}_{pc} + \mathcal{H}_c$, where $\mathcal{H}_p$ is the structure-based potential of protein, $\mathcal{H}_{pc}$ is the potential between protein and crowders, and $\mathcal{H}_c$ is the potential between crowders. Every residue is represented by a single unit and is strung together into a polymer chain. The potential consists of attractive native interactions and repulsive nonnative interactions, as well as the backbone geometry by means of bond and angle interactions. To include the desolvation effect at different pressures, we implemented a pressure-dependent desolvation potential [9,11,12], which contains a *desolvation barrier* and a *water-separated minimum* at 0.5 σ and 0.8 σ from the contact minima, respectively, which is about the diameter of a water molecule from the native potential well (σ is the reduced unit equivalent to 3.8Å). The Hamiltonian of this structure-based model is as follows:

$$\begin{aligned}\mathcal{H}_p(\Gamma,\Gamma^0) &= \sum_{i<j} K_r\,(r_{ij}-r_{ij}^0)^2\delta_{j,i+1} + \sum_{i\in\text{angles}} K_\theta\,(\theta_i-\theta_i^0)^2 \\ &+ \sum_{i\in\text{dihedrals}} K_\phi\left(\{1-\cos[\phi_i-\phi_i^0]\}+\frac{1}{2}\{1-\cos[3(\phi_i-\phi_i^0)]\}\right) \\ &+ \sum_{\substack{i,j\in\text{native}\\ |i-j|>4}} U(r_{ij}) + \sum_{i,j\notin\text{native}} \epsilon\left(\frac{\sigma}{r_{ij}}\right)^{12}, \qquad \text{(S1)}\end{aligned}$$

where Γ is a configuration of the set $r, \theta, \phi$. The $r_{ij}$ term is the distance between $i^{th}$ and $j^{th}$ residues, $\theta$ is the angle between three consecutive beads, and $\phi$ is the dihedral angle defined over four sequential residues. $\delta$ is the Kronecker delta function. The native state values of $r^0, \theta^0, \phi^0$ were obtained from their crystal structure configuration, $\Gamma^0$, which $= \{\{r^0\},\{\theta^0\},\{\phi^0\}\}$. In the backbone terms, $K_r$, $K_\theta$, and $K_\phi$ are force constants of the bond, bond-angle, and dihedral potentials, respectively. We used $K_r = 100\epsilon$, $K_\theta = 20\epsilon$, and $K_\phi = \epsilon$, where $\epsilon$ is the solvent averaged energy ($\epsilon$ =0.6 kcal/mol). In the native contacts term, the desolvation potential, $U(r)$, is shown in Eqn. S2 and in Fig. S2.2. This potential has been modified to incorporate high pressure in which the solvent averaged energy $\epsilon$ becomes less than zero (second term of Eqn S2).

$$U(r) = \begin{cases} \epsilon Z(r)(Z(r)-2) & \text{with } Z(r) = \left(\frac{r'}{r}\right)^k \quad \text{if } r < r' \text{ and } \epsilon > 0 \\ -\epsilon Z(r)(Z(r)-2) - 2\epsilon & \text{with } Z(r) = \left(\frac{r'}{r}\right)^k \quad \text{if } r < r' \text{ and } \epsilon < 0 \\ CY(r)^n \frac{\frac{Y(r)^n}{2} - (r^\dagger - r')^{2n}}{2n} + \epsilon'' & \text{with } \begin{cases} Y(r) = (r - r^\dagger)^2 \\ C = \frac{4n(\epsilon + \epsilon'')}{(r^\dagger - r')^{4n}} \end{cases} \quad \text{if } r' \le r < r^\dagger \\ -B\frac{Y(r) - h_1}{Y(r)^m + h_2} & \text{with } \begin{cases} Y(r) = (r - r^\dagger)^2 \\ B = \epsilon' m (r'' - r^\dagger)^{2(m-1)} \\ h_1 = \frac{\left(1 - \frac{1}{m}\right)(r'' - r^\dagger)^2}{\frac{\epsilon'}{\epsilon''} + 1} \\ h_2 = \frac{(m-1)(r'' - r^\dagger)^{2m}}{\frac{\epsilon''}{\epsilon'} + 1} \end{cases} \quad \text{if } r^\dagger \le r \end{cases} \tag{S2}$$

The desolvation potential is a function of the distance, $r$, between residues in the native contact pairs with $r'$ as the minimum of the first potential well, $r^\dagger$ as the maximum of the desolvation barrier, and $r''$ as the minimum of the second potential well. The separation between $r'$ and $r''$ is the size of a single water molecule, 0.8 σ. The terms $r'$, $r''$, and $r^\dagger$ satisfy such relation: $r^\dagger = (r' + r'')/2$. We used $m = 3$, $k = 6$, $n = 1$ for the potential constants. The solvent average energy, $\epsilon$ is the depth of the first well, $\epsilon'$ is the depth of the second well, $\epsilon''$ is the height of the desolvation barrier. The value of $\epsilon$, $\epsilon'$, and $\epsilon''$ is related to the magnitude of pressure, $P$, by the following equation [12] (Eqn S3).

$$\begin{cases} \epsilon = 0.6 - 0.076 \cdot P \\ \epsilon'' = 0.8 + 0.127 \cdot P \\ \epsilon' = 0.2 \end{cases} \tag{S3}$$

where the unit of $\epsilon$, $\epsilon'$, and $\epsilon''$ is kcal/mol, and the unit of pressure is dimensionless (where $\epsilon/\sigma^3 = 760$ bar).

Lastly, the potentials $\mathcal{H}_c$ and $\mathcal{H}_{pc}$ are as follows:

$$\mathcal{H}_c = \sum_{i>j}^{n_c} \epsilon \left(\frac{\sigma_{ij}}{r_{ij}}\right)^{12} \tag{S4}$$

$$\mathcal{H}_{pc} = \sum_{i}^{N} \sum_{j}^{n_c} \epsilon \left(\frac{\sigma_{ij}}{r_{ij}}\right)^{12} \tag{S5}$$

where $\sigma_{ij}$ is the distance between two particles in contact (either residue and crowder, or crowder and crowder), given by $\sigma_{ij} = 0.5(\sigma_{ii} + \sigma_{jj})$. $N$ is the number of residues, which is 415, and $n_c$ is the number of crowders. To achieve a crowder volume fraction of $\phi$ = 40%, $n_c$ = 7222. This is from $\phi = n_c \times \frac{4}{3}\pi\sigma_c^3/V_{box}$, where $V_{box}$ is the volume of the cubic box (length = 600σ), and $\sigma_c = 2.3R_{gyr}^0$ or 55 Å. $\sigma_c$ is the radius of Ficoll 70 [13,14]. ($R_{gyr}^0$ is the radius of gyration of PGK crystal structure $R^0_{gyr}$ = 23.6 Å).

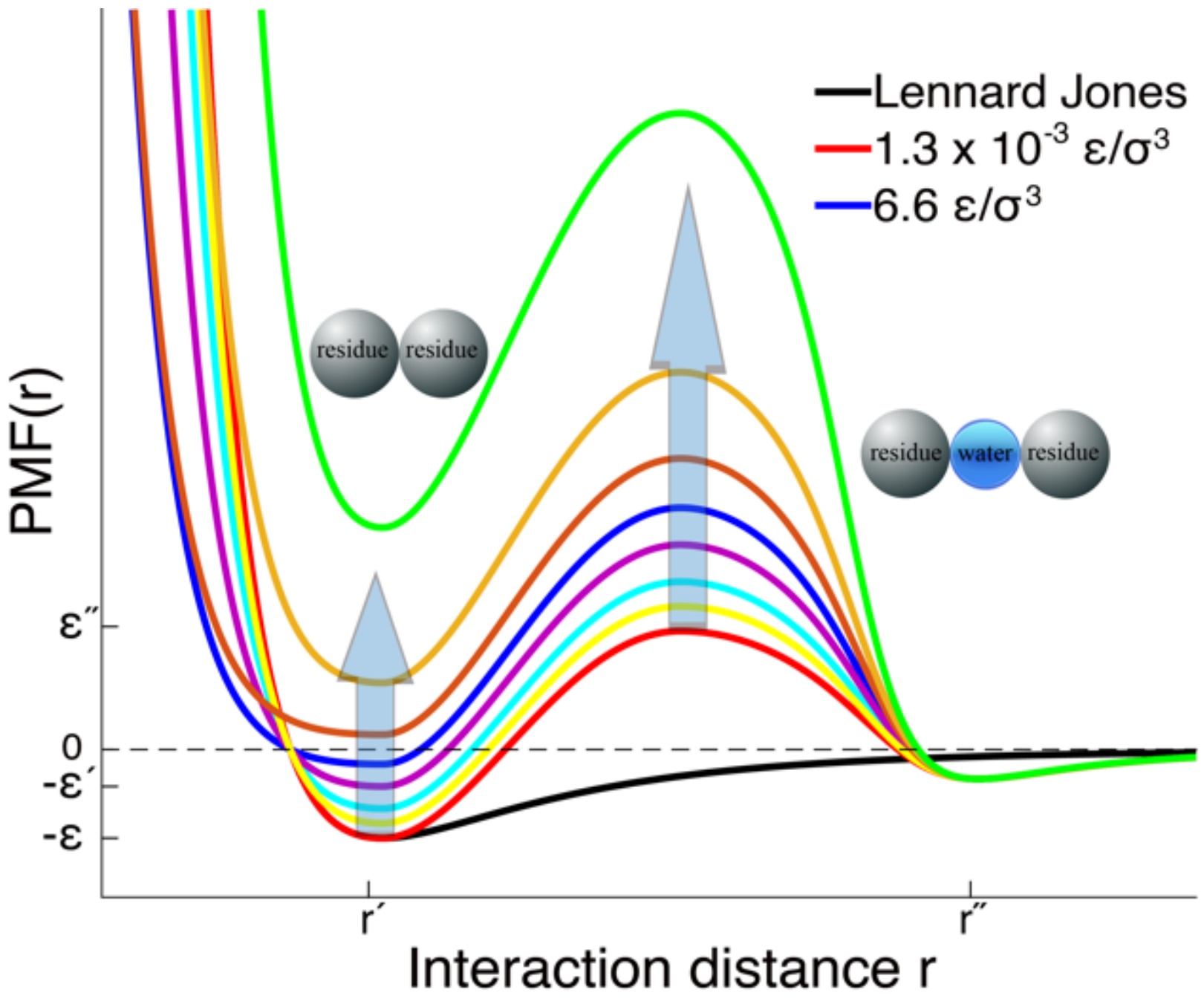

**Figure S2.2** A schematic representation of a phenomenological potential for tertiary contact formation, which includes the possibility of desolvation. $r'$ and $r''$ label the residue-residue contact minimum and the single water molecule-separated contact minimum, respectively. Three regions are defined: (i) when the separation distance between residues r is shorter than the range of the desolvation barrier, then a native contact is formed; (ii) when the distance between residues r is larger than the 1.5 times the water molecule distance (1.2 σ), no contact is formed. (iii) Otherwise, the residues are separated by about the size of a single water molecule, then a "water mediated contact" is formed; When the pressure increases, both the energy of the residue-residue contact and the energy of the desolvation potential shift upward as shown by arrows in the figure. $\sigma^3 P/\varepsilon$ = 1.3 x $10^{-3}$, 1.3, 4.6, 6.6, 9.2, 13.8, 18.4, and 27.6 are represented by the red, yellow, cyan, purple, blue, brown, orange, and green colored lines, respectively. The Lennard-Jones potential is plotted in black for comparison.

## S2.3 Simulation details

We performed all simulations using GROMACS [15] to integrate Langevin equations of motion at a low friction limit. We used SMOG: Structure-based Models for Biomolecules [16] software (http://smog-server.org/) to prepare the model for GROMACS simulations. The natural time unit of a coarse-grained model is $\tau_L = (m\sigma^2/\varepsilon)^{1/2}$ (approximately 1.9 ps), where $m$ is the mass of the $C_\alpha$ bead (at 100a.m.u), σ is the unit length (equivalent to 3.8Å, the bonding distance between two adjacent $C_\alpha$ beads), and ε is the solvent mediated interaction (0.6 kcal/mol). The integration time step is $10^{-3}$ $\tau_L$.

The protein systems with no crowders were equilibrated for $3.6 \times 10^4$ $\tau_L$ over a wide range of temperatures. We then employed a highly scalable technique of the Replica Exchange Method (REM) [17,18] for enhanced sampling of the simulations with the initial structures taken from the equilibration step. We set up 40 replicas over a temperature range of $0.67 < k_BT/\varepsilon < 1.53$. An exchange between neighboring replicas was attempted at every 75 $\tau_L$ [19]. We set up pressures at $\sigma^3P/\varepsilon$ = 1.3 x $10^{-3}$, 1.3, 2.6, 4.6, 6.6, 9.2, 13.8, 18.4, 23 and 27.6. The trajectories we sampled at every 1.5$\tau_L$ and the number of samples for each temperature ranges from 200K to 1 million, depending on the convergence of the data. A system with a Lennard-Jones potential was simulated and used as a control.

In the presence of crowders ($\phi$ =20% and 40%), we sampled every 2$\tau_L$, and equilibrated for $10^2$ $\tau_L$ over the same temperatures used at $\phi$ = 0%. REM was also used with the same details used at $\phi$ = 0%. We set up pressures at $\sigma^3P/\varepsilon$ = 1.3 x $10^{-3}$, 4.6, 6.6, 9.2, and 18.4.

### S2.4 Order parameters

**Shape, asphericity, and radius of gyration** To determine the shape of a given conformation, we used a radius of gyration $R_{gyr}$, and two rotationally invariant quantities: shape parameter, $S$, and asphericity parameter, Δ. These quantities are calculated with the use of the inertia tensor, $T$ [20]:

$$T_{\alpha\beta} = \frac{1}{2N^2}\sum_{i,j=1}^{N}(r_{i\alpha} - r_{j\alpha})\cdot(r_{i\beta} - r_{j\beta}) \qquad \text{(S6)}$$

where $N$ is the number of residues in the protein, $r_{i\alpha}$ is the position of bead $i$ and α, β are the coordinates $x$, $y$, $z$. The eigenvalues of $T$ and $\lambda_i$ are the squares of the three principal radii of gyration. Therefore,

$$R_{gyr}^2 = tr(T) = \sum_{i=1}^{3}\lambda_i. \qquad \text{(S7)}$$

Asphericity Δ is calculated by using

$$\Delta = \frac{3\sum_{i=1}^{3}(\lambda_i - \bar{\lambda})^2}{2tr(T)^2} \qquad \text{(S8)}$$

where $\bar{\lambda} = tr(T)/3$. Lastly, the shape parameter, $S$, is calculated by

$$S = \frac{27\prod_{i=1}^{3}(\lambda_i - \bar{\lambda})}{tr(T)^3} \qquad \text{(S9)}$$

The parameters $S$ and Δ are in the ranges $0 \le \Delta \le 1$ and $-0.25 \le S \le 2$. For a perfect sphere, $S = \Delta = 0$. Δ greater than 0 is an indication of the extent of anisotropy. Negative and positive values of $S$ refer to oblate and prolate ellipsoids respectively.

**Overlap function, native contact formation, water-mediated native contact formation** The overlap function $\chi$ [21], fraction of native contacts $Q$, fraction of native contacts for N-domain $Q_N$, and C-domain $Q_C$, and fraction of native water-mediated contacts *pseudo Q* [9] are used to measure the similarity to the crystal structure.

The overlap function $\chi$ is defined as

$$\chi = 1 - \frac{1}{N^2 - 5N + 6}\sum_{i=1}^{N-3}\sum_{j=i+3}^{N}\Theta\left(1.2r_{ij}^0 - r_{ij}\right) \qquad \text{(S10)}$$

where Θ is the Heaviside step-function, $r_{ij}$ is the distance between the beads $i$ and $j$ for a given conformation, and $r_{ij}^0$ is the corresponding distance in the crystal structure. $\chi$ = 0 means most similar and =1 means most dissimilar to crystal structure. Note that $\chi$ includes almost all inter-residue distance pairs, not just the native contact pairs as defined for $Q$.

$Q$ is the fraction of the native contact formation and is defined as

$$Q = \frac{1}{\mathcal{N}} \sum_{i,j \in \text{native}} \Theta\left(\delta - \left|r_{ij} - r_{ij}^{0}\right|\right) \tag{S11}$$

where $\mathcal{N}$ is the number of native contacts, and $\delta$ is a cutoff value = 0.4 σ. $Q_N$ $(Q_C)$ is the fraction of native contacts for N-domain (C-domain). *pseudo Q* is the fraction of water-mediated native contact formation.

$$pseudo\ Q = \frac{1}{\mathcal{N}} \sum_{i,j \in \text{native}} \Theta\left(\left|r_{ij} - r_{ij}^{0}\right| - \delta\right) - \Theta\left(\left|r_{ij} - r_{ij}^{0}\right| - \delta'\right) \tag{S12}$$

where $\delta'$ is a cutoff value = 1.2 σ. $Q_N$, $Q_C$, and *pseudo Q* are in the range from 0 (no contacts) to 1 (maximum number of contacts).

To measure the orientation of the N-domain relative to the C-domains of PGK we use $\cos(\theta)$ (previously used for characterizing the state of PGK in Ficoll 70 in ref [4]). This quantity, $\cos(\theta)$, is equal to the inner product of two vectors, one vector from each domain. Residues from 50 to 74 and residues from 287 to 360 define the two vectors. The two vectors are parallel in the crystal structure[4], which gives $\cos(\theta) = 1$.

### S2.5 Volume calculations

Before computing the cavity volume or co-volume of a protein conformation, we reconstructed the all-atomistic protein structures using the "Side-chain $C_\alpha$ to all-atom" (SCAAL) [22] program.

**Co-volume** We used the volume calculator, 3v, [23] to compute the co-volume [24]. This is done by rolling a hard sphere over the surface of a protein conformation. The hard sphere has a radius equal to the crowder (Ficoll 70) radius, $\sigma_c$, which is 55 Å [13,14].

**Cavity volume** A similar software, McVol [25], was used to compute the conformation's cavity volumes, using a probe size of 1.4 Å, the radius of a water molecule.

## S3. PGK's $\phi - P$ phase diagram from computer simulations

### S3.1 Additional evidence for schematic phase diagram

The crowder volume fraction-pressure ($\phi - P$) schematic phase diagram in Fig. 2(a) and 2(c) were constructed from the information provided by pressure melting curves of isothermal ensembles with respect the overlap function $\chi$, and free energy landscapes with respect to order parameters radius of gyration $R_{gyr}$ and $\chi$ at various temperatures and pressures, including Fig. 2(b) and 2(d). Fig. S3.1 illustrates the folding/unfolding transition at $\phi$ = 0, 0.2, and 0.4, plotting the ensemble average $\langle\chi\rangle$ as a function of pressure at varying temperatures (top), and as a function of temperature at varying pressures (bottom). At high temperature, PGK jumps from a folded ($\chi$ = 0) to unfolded state ($\chi$ = 0.9) over a short pressure increment reflecting its two-state behavior. At lower temperature ranges, a multi-state transition occurs with the melting pressure at 6.6 $\varepsilon/\sigma^3$. These lower range temperature systems contain multiple intermediate states as pressure is increased from 2.6 to 7.8 $\varepsilon/\sigma^3$. Note that intermediates can be found in a range of temperatures and pressures close to these conditions.

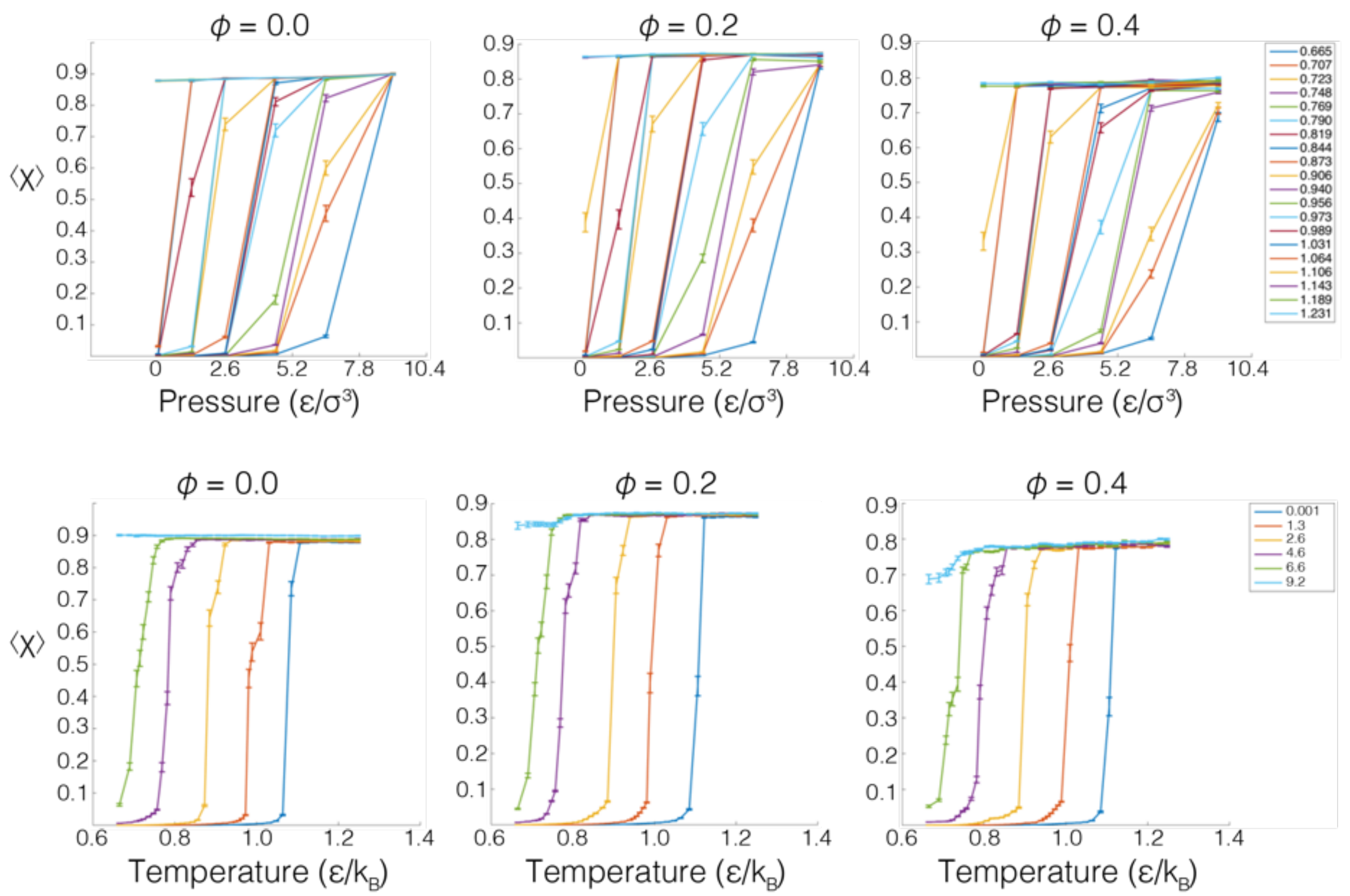

**Figure S3.1** Average Overlap as a function of pressure (top) and temperature (bottom) at $\phi$ = 0%, $\phi$ = 20%, and $\phi$ = 40% from left to right. The legend on the top row indicates temperatures $\varepsilon T/k_B$, and the legend on the bottom row indicates pressures $\varepsilon P/\sigma^3$. The error bars error bars are calculated via jackknife method.

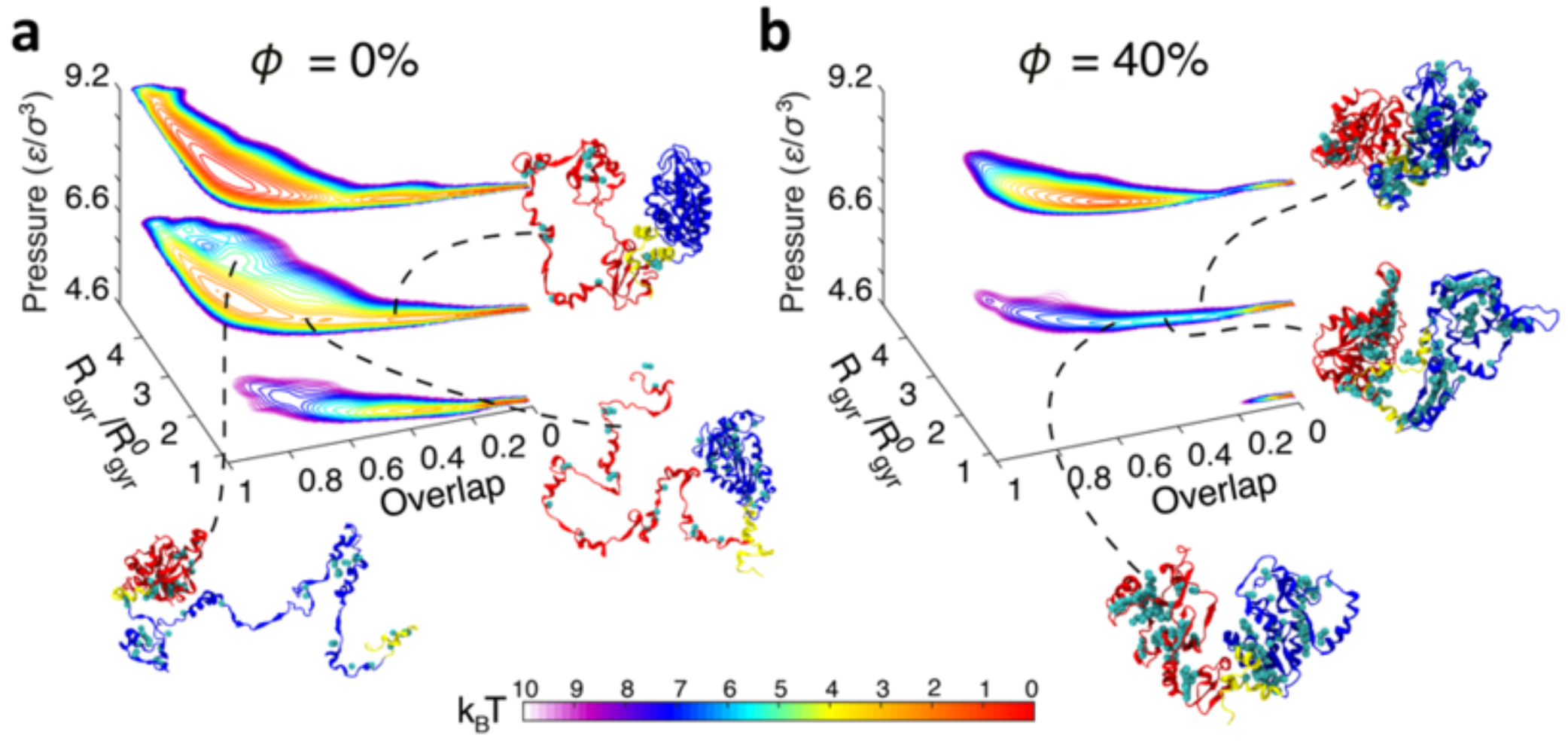

**Figure S3.2 Free energy landscape and impact of crowding on intermediate.** (a-b) Free Energy Landscapes with respect to the radius of gyration $R_{gyr}/R^0_{gyr}$ and overlap function for pressures 4.6, 6.6, and 9.2 $\varepsilon/\sigma^3$ at $k_BT/\varepsilon$ = 0.97 without crowders (a) and with crowders (b). The fraction of native contacts is normalized such that the completely unfolded state is zero, and the folded state is one. The color of the contour is scaled by $k_B$T. Structures represent 6.6 $\varepsilon/\sigma^3$ ensembles that fall between $Q$ = 0.2 to 0.7 in the absence and presence of crowding agent. N-, C-domain, and hinge are in red, blue, and yellow respectively. A cyan sphere was inserted in between residues to show water-mediated contacts.

## S3.2 Characterization of PGK's structures in each phase

We use several order parameters (described in S2.4) to characterize the structures of PGK (Table S3.1). The crystal state (C) is the folded state most similar to the PDB crystal structure with $0.94 \leq R_{gyr}/R^0_{gyr} \leq 1.02$, $S = 0.27$ and $\Delta = 0.27$. The spherical state (Sph) is a collapsed state with torsion on the hinge with $0.9 \leq R_{gyr}/R^0_{gyr} \leq 1.2$, $S = 0.1$ and $\Delta = 0.1$. Water-mediate contacts are formed on the backside of both domains and hinge. The intermediate (I) contains an unfolded N-domain and folded C-domain, appearing at pressure of 6.6 $\varepsilon/\sigma^3$ and temperature of 0.97 $\varepsilon/k_B$. Without crowders to restrict the available volume, water-mediate contacts are not stable and unravel that region of the protein. The water-swollen unfolded ensemble (SU) are configurations that have 30% to 40% water-mediated contacts and other contacts are unfold with $1.5 \leq R_{gyr}/R^0_{gyr} \leq 1.6$. In crowded conditions, water-mediated contacts are more dominant than a completely unfolded contact. The unfolded state (U) is completely unfolded with less than 10% water-mediated contacts with $2 \leq R_{gyr}/R^0_{gyr} \leq 2.7$.

**Table S3.1** Characteristics of PGK's representative structures from each phase

| | **Crystal (C)** | **Collapsed Crystal (CC)** | **Spherical (Sph)** | **Intermediate (I)** | **Water-swollen unfolded (SU)** | **Unfolded (U)** |
|---|---|---|---|---|---|---|
| $R_{gyr}/R^0_{gyr}$ | 0.94 to 1.02 | 0.9 to 1.2 | 0.9 to 1.2 | 1.31 to 1.62 | 1.5 to 1.6 | 2.0 to 2.7 |
| $\chi$ | 0 | 0.3 to 0.4 | 0.3 to 0.4 | 0.4 to 0.6 | 0.6 to 0.7 | 0.9 |
| $S$ | 0.27 | 0.2 | 0.1 | 0.34 | 0.27 | 0.48 |
| $\Delta$ | 0.27 | 0.17 | 0.1 | 0.31 | 0.29 | 0.39 |
| $Q_N$ | 1 | 0.8 | 0.8 | 0.1 | 0.07 | 0 |
| $Q_C$ | 1 | 0.6 | 0.6 | 0.89 | 0.06 | 0 |
| *pseudo Q* | 0 | 0.15 | 0.15 | 0.06 | 0.35 | 0.08 |
| $Cos(\theta)$ | 1 | 0.5 to 0.75 | -1 to -0.5 | n/a | n/a | n/a |

To distinguish the various configuration states of PGK, we used a k-means clustering algorithm [26,27] to identify states under high pressure ($\sigma^3 P/\varepsilon = 6.6$ and $k_B T/\varepsilon = 0.97$). We focused on clustering structures within the range of fraction of native contacts $Q$ = 0.2 to 0.7, which excludes the crystal state structures and completely unfolded structures. A vector, $\boldsymbol{X}$, with elements $x_j = [x_{1j}, x_{2j}, \ldots x_{Pj}]$, characterizes each cluster. For the purposes of this study, the elements of $\boldsymbol{X}$ are the order parameters as shown above. Once clustered, the representative coarse-grained intermediate structures are naturally selected as the center of the most populated clusters. These are the structures shown in Fig 2 and S3.2.

## S4. PGK's P – T phase diagram from fluorescence measurements with and without Ficoll

We explored the pressure-temperature phase diagram of phosphoglycerate kinase using fluorescence spectroscopy. Fig. S4.1a shows a series of fluorescence spectra that were acquired at different pressures from 0.1 MPa to 250 MPa and Fig. S4.1b shows the same in the presence of 200mg/ml Ficoll 70. The data were analyzed by mean wavelength shift in the main text. Here we discuss an alternative analysis in terms of singular value decomposition (SVD) that also supports the two- and three state assignments shown in Fig. S2.1.

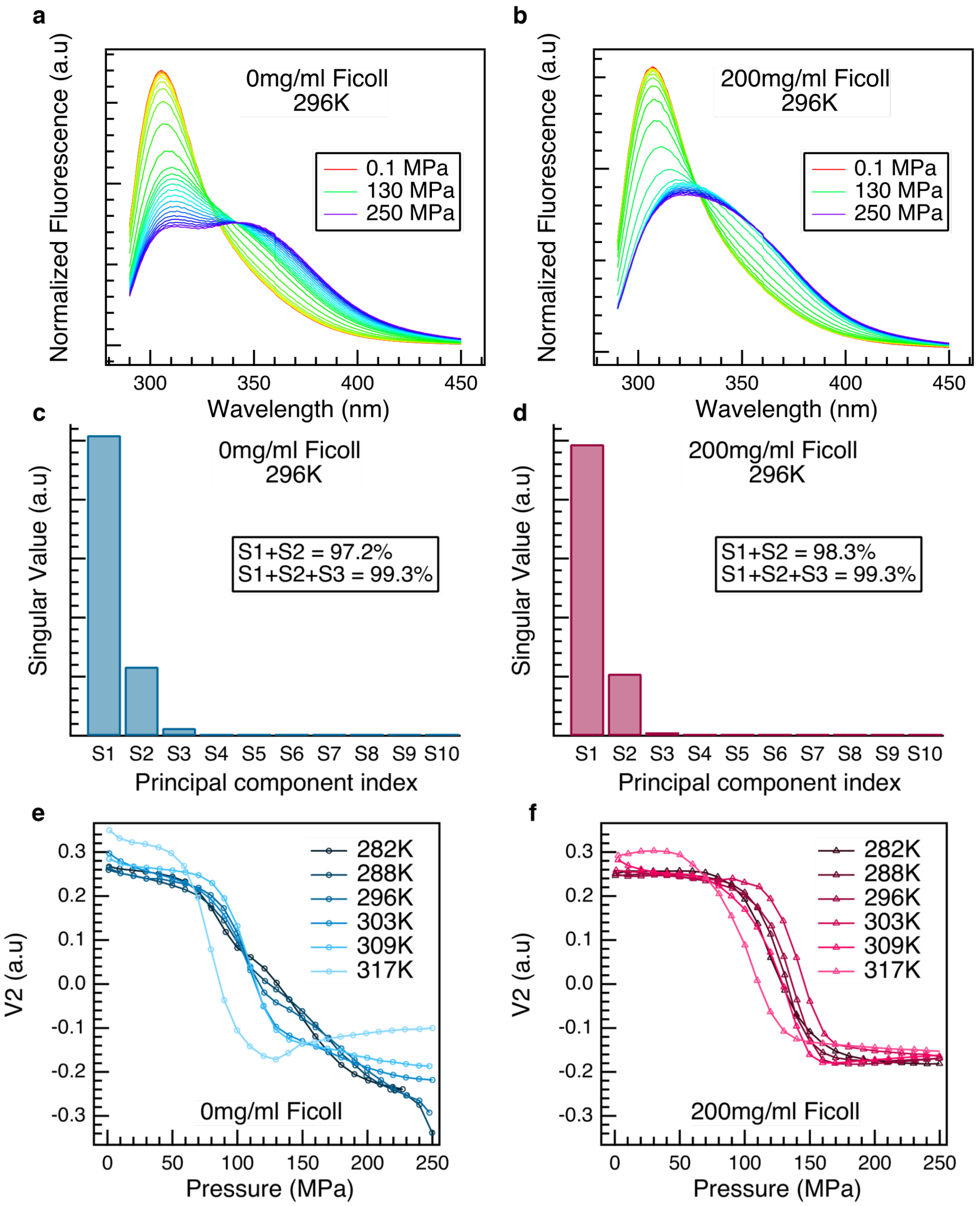

**Figure S4.1** Using Singular value decomposition on raw spectra to confirm three state to apparent two state transition. Raw fluorescence spectra of PGK taken in increments of 10 MPa at 296K (23 °C) from 0.1 MPa to 250 MPa. (a) in the absence and (b) presence of 200mg/ml Ficoll 70. (c) and (d) The singular values for SVD of the data shown in (a) and (b) respectively. As can be seen the third component is more significant in the absence of Ficoll 70 supporting an apparent three state transition, while in the presence of Ficoll 70 two components seem enough. (e) The second component in the pressure space (V2) plotted for different temperatures. Till 303K PGK seems to go through a quasi-three-state transition, while at 309K and 317K it seems to be an apparent two-state transition. The rise in the component at 317K near 100 MPa seems to be occurring due to hyperfluorescence of Tryptophan. This was seen at 317K for all concentrations of Ficoll 70 except 200mg/ml. (f) The second component in the pressure space (V2) plotted for different temperatures in the presence of 200 mg/ml Ficoll 70. At all temperatures an apparent two-state model can be seen showing that increased Ficoll 70 concentration leads to the disappearance of an intermediate state.

**Singular value decomposition of fluorescence melts** Tryptophan is known to hyperfluoresce near protein unfolding transitions. Hyperfluorescence is an increase and then decrease of fluorescence intensity during the transition caused by increased mobility of the tryptophan (reduces quenching in the native structure), and subsequent lower fluorescence once solvent-exposed [29]. Thus, the fluorescence data in Fig. S4.1ab is normalized to the same integrated fluorescence intensity for comparison. We performed singular value decomposition on the raw fluorescence data. The first SVD component roughly follows the overall intensity of the average fluorescence data, and is not of interest here. The amplitude V2 as a function of pressure of second SVD component (Fig. S4.1ef) follows the trend in the pressure melt. Note the double transition (apparent three states) between 282K to 303K in the absence of Ficoll 70 (Fig. S4.1e) whereas a single transition (apparent two-state equilibrium) is observed in the components at 309 K and 317 K. At 317 K, there is a small dip due to hyperfluorescence in the main transition at 120 MPa. This dip was not observed at the highest concentration of Ficoll 70 (200 mg/ml) in Fig. S4.1.f, suggesting that increased Ficoll 70 concentration restricts the mobility of the tryptophan. SVD analysis in 200 mg/ml Ficoll 70 shows no evidence of a double transition at any temperature, so the third state (intermediate) has disappeared. Note also that the third singular value without Ficoll 70 (Fig. S4.1c) is roughly twice as large as the third singular value in 200 mg/ml Ficoll 70, also indicating that high Ficoll 70 concentration conditions approach two-state behavior more closely.

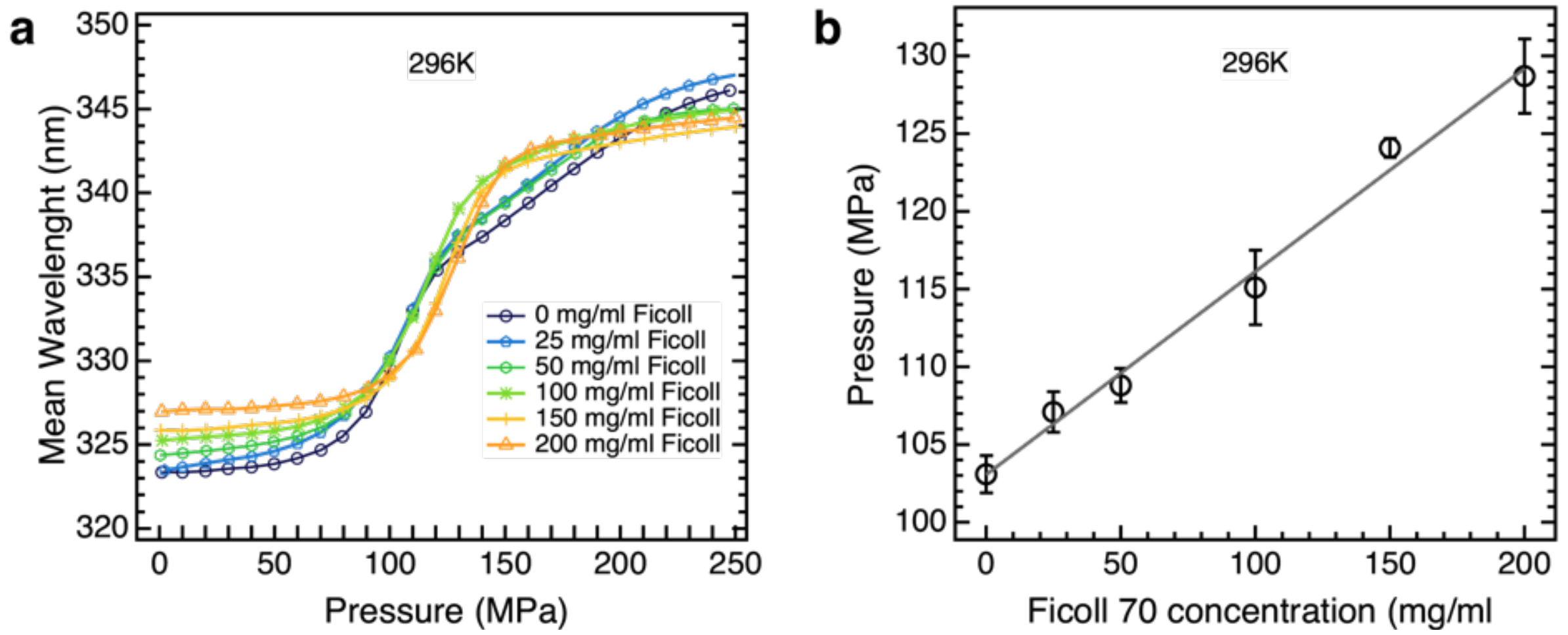

**Figure S4.2** Pressure-induced denaturation of phosphoglycerate kinase at various Ficoll 70 concentrations. (a) Fluorescence pressure melts plotted at different concentration of Ficoll 70 and at 296K. (b) Midpoint pressures obtained from the fits to data shown in panel (a) plotted as a function of Ficoll 70 concentration. Only midpoint pressures of transition 1 are plotted. The solid line is a linear fit of the data.

Representative pressure denaturation curves of PGK at 296 K and various Ficoll 70 concentrations are shown in Fig. S4.2, analyzed by mean wavelength. We found again that a double transition (apparent three states, i.e. an intermediate) is observed without Ficoll 70, but at 150 and 200 mg/mL Ficoll 70 only a two-state transition occurs. The transition pressures for these conditions along with the first transition pressure of the sample without Ficoll 70 are plotted in Fig. S4.2a. There is a linear increase in pressure stability with increasing Ficoll 70 concentration (Fig S4.2b).

The pressure-temperature phase diagram of PGK was constructed using the equilibrium denaturation data with and without Ficoll 70. A transition from an intermediate at 23 °C without Ficoll 70 to a two-state pressure-induced transition with 100 mg/mL Ficoll 70 is shown in Fig. S4.3 (green). At 15 °C, an intermediate was observed in both cases (gray), while at 30 °C, a two-state transition was observed in both cases (red).

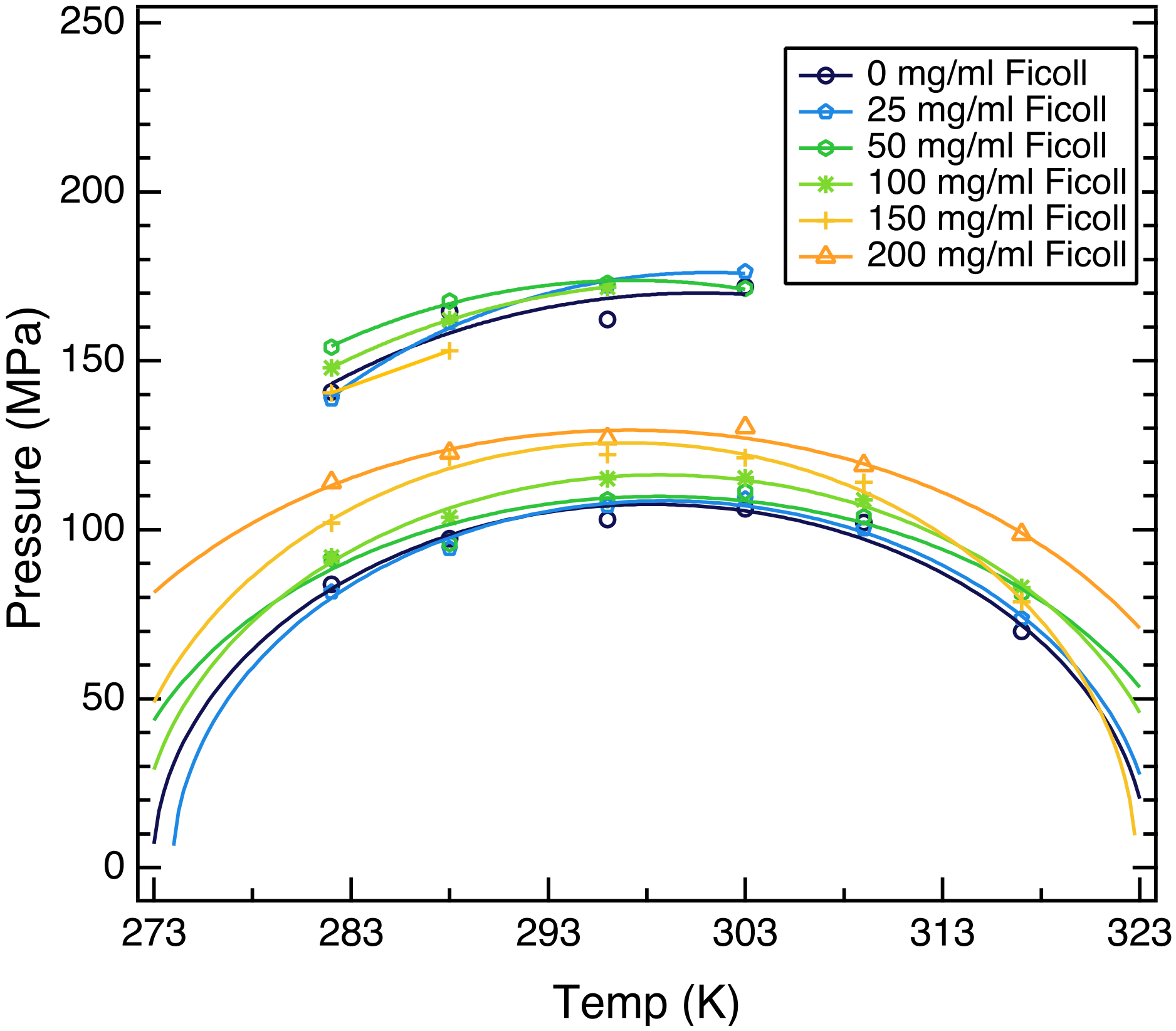

**Figure S4.3** Experimental $P$-$T$-$\phi$ phase diagram of PGK at all concentrations of Ficoll 70. The second transition is not observed at higher temperatures. At [Ficoll 70]=200 mg/ml, only one transition is observed at all temperatures. Solid elliptical curves going through the circles are fits representing the $\Delta G = 0$ curves fit with Hawley's equation. The shrinking of the second transition can be noticed as the Ficoll 70 concentration increases. Error bars for each fit point are not shown to maintain clarity.

**Fluorescence melts of the N-Terminal domain of PGK** We performed fluorescence melts on just the N-terminal domain of PGK to assess how stable the N-terminal domain, predicted by our theory to unfold first in the intermediate state, is. As can be seen from Fig. S4.4 the N-terminal domain starts out unfolded and remains unfolded in the presence of Ficoll 70. The CD spectra (not shown) indicates the presence of more alpha helix instead of beta sheets as expected in the N-terminal domain of PGK. This shows that the N-terminal domain is precarious to unfolding and might be producing one of the two transitions in the apparent three state model.

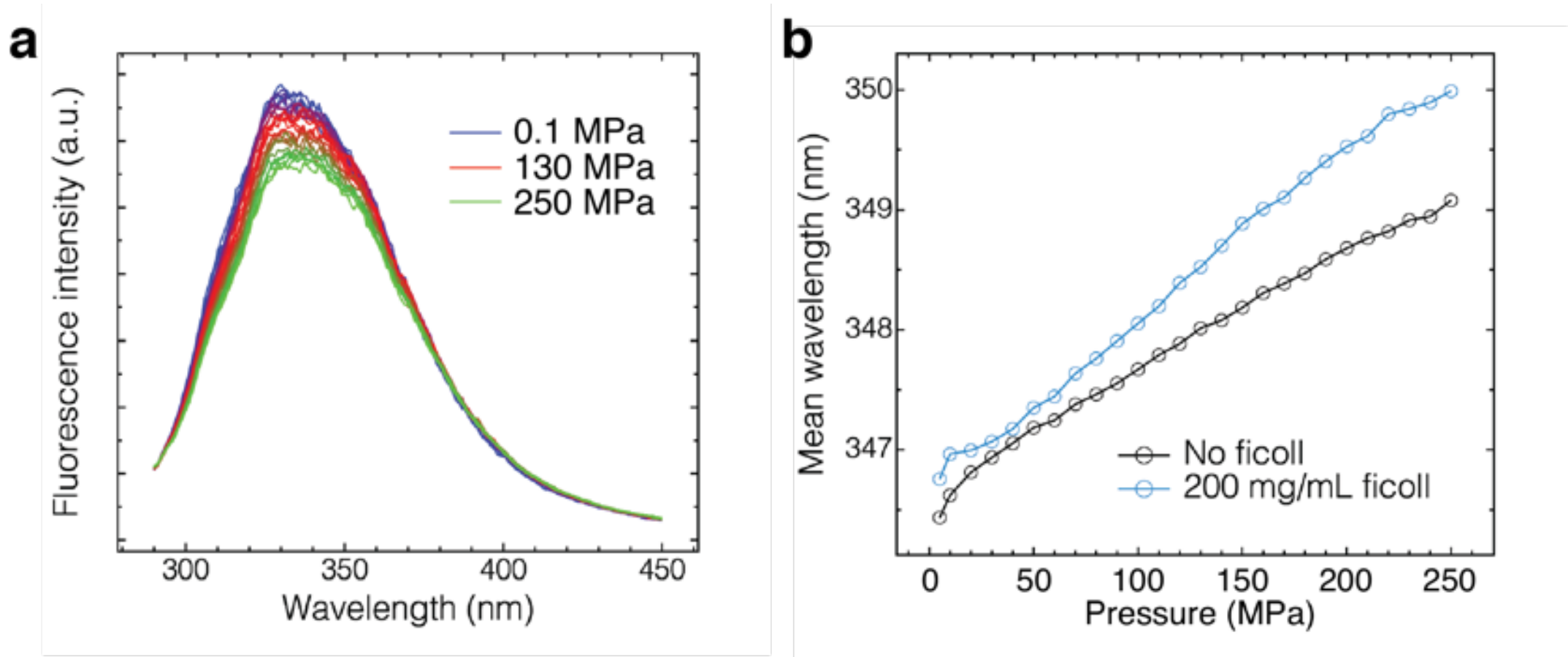

**Figure S4.4** (a) Change in the fluorescence spectra of the N-terminal domain of PGK as the pressure is increased from 0.1 MPa to 250 MPa in steps of 10 MPa. (b) The N-terminal domain is unfolded in the absence or presence of Ficoll 70 showing that N-terminal domain can unfold easily.

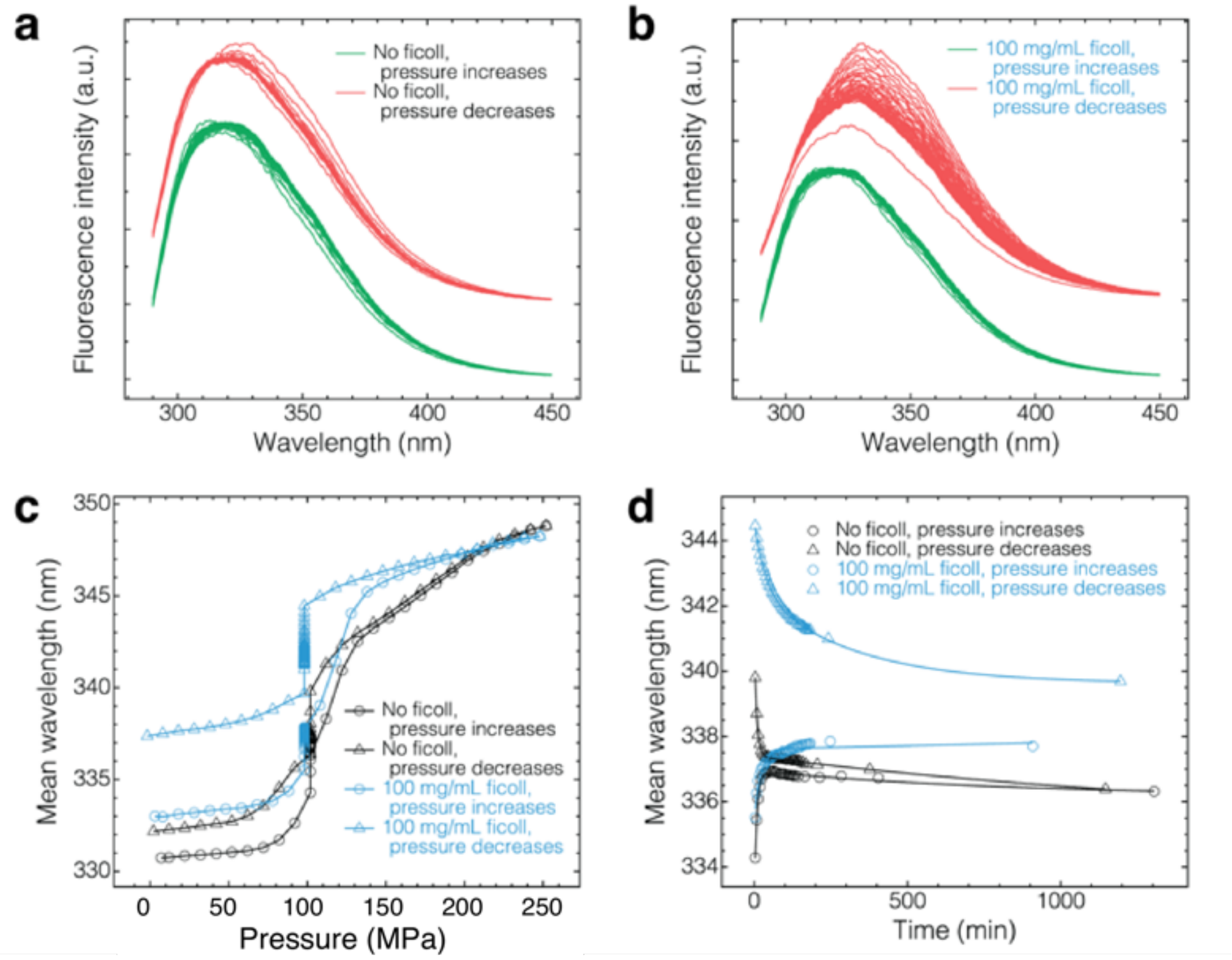

**Figure S4.5** Hysteresis in the pressure thermodynamics of PGK. (a) Change in fluorescence spectra as the pressure is increased and decreased at 100 MPa without Ficoll 70. (b) Change in fluorescence spectra as the pressure is increased and decreased at 100 MPa in 100 mg/mL Ficoll 70. (c) Hysteresis in the pressure thermodynamics of PGK with (blue) and without (black) Ficoll 70 as the pressure is increased (circles) and as the pressure is decreased (triangles). (d) Kinetics at 100 MPa as the pressure is increased (circles) and as the pressure is decreased (triangles) with (blue) and without (black) Ficoll 70.

**Reversibility and hysteresis** Pressure denaturation was done in a series of 10 MPa steps. After each step, the system was allowed to equilibrate for 3 minutes and a spectrum was taken. The pressure was then further increased by 10 MPa and the procedure was repeated. However, some hysteresis was observed, indicating that high-pressure kinetics of PGK transitions is slower than 3 min. This hysteresis is quantified in Figure S4.5. We increased the pressure using the 3-minute intervals up to 100 MPa and let the system equilibrate while monitoring fluorescence spectra. We repeated the same procedure when the pressure was decreased back to 0.1 MPa from the final pressure of 250 MPa and in the presence of 100 mg/mL Ficoll 70. When the pressure is increased, hysteresis will cause overestimation of the unfolding pressure (overestimation of stability) but the number of transitions (and states) would not be affected.

**S5. Pressure-induced unfolding mechanism in the presence and absence of crowding**
In Fig S5.1, we explore the impact of pressure and crowding on the folding of PGK in more detail through the distributions of the cavity volume (conjugate of pressure) and co-volume (the conjugate of crowding volume fraction; the excluded volume of the protein with respect to the crowder). There is a clear entropic trade off in Fig S5.1(a) and S5.1(b). As pressure increases, cavity volume decreases (Fig. S5.1(a)) and co-volume increases (Fig. S5.1(b)) due to unfolding. PGK remains in either a Sph or a CC state (decreasing co-volume) at $\phi = 40\%$, even after it initially cracks at 6.6 $\varepsilon/\sigma^3$ resulting in an increase in cavity volume. Comparison of $\phi = 0\%$ (blue) with $\phi = 40\%$ (orange) in Fig. S5.1(a) and S5.1(b) show that macromolecular crowding favors collapsed states with higher cavity volumes and lower co-volumes, signifying an increase in stability of folded structures with the increase of crowding agent. Furthermore, the protein's unfolded state is less populated, which is in agreement with the upward shift in the blue curve of the *P-T* phase diagram in Fig S5.1(a). As $\phi$ increases, the critical line is passed. This is shown in figure S5.1(b) by the probability distribution of three populations that merge into a two, signifying the coalescence of states U and I, as also observed experimentally.

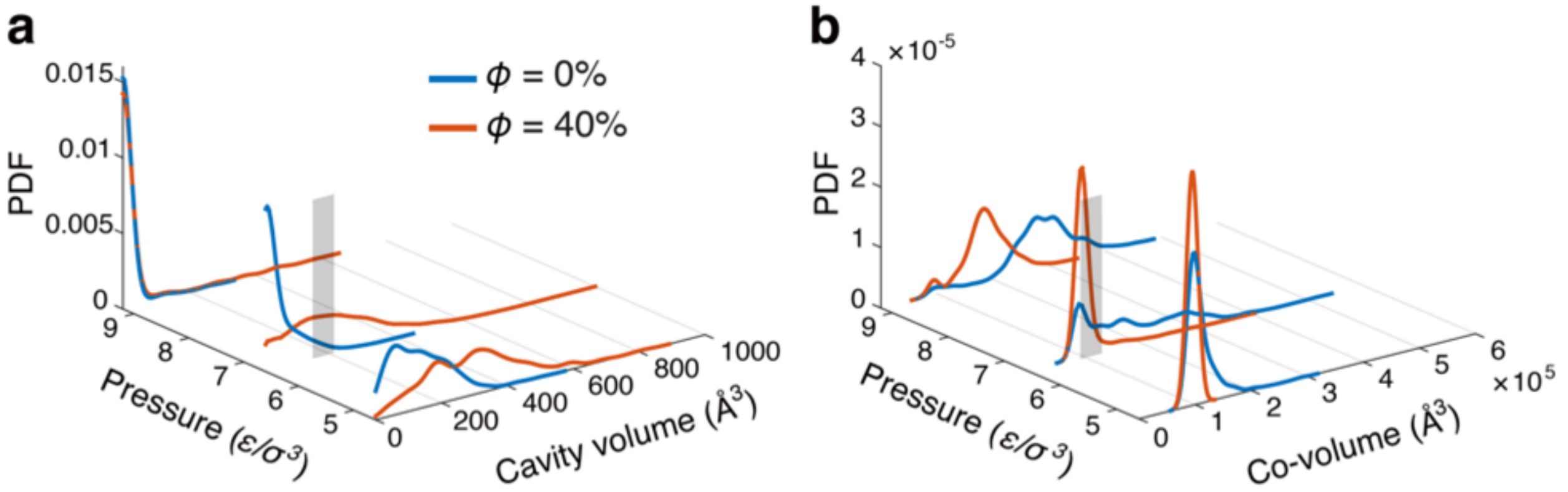

**Figure S5.1** Volume distributions and structures near critical regime. Cavity volume (a) and co-volume (b) distributions for PGK at $k_BT/\varepsilon$ = 0.97 and pressures 4.6, 6.6, and 9.2 $\varepsilon/\sigma^3$ with (orange) and without (blue) crowding. The conformations from the ensemble seen in Fig 5C of the manuscript are shown by the grey shaded regions in a and b.

**Three-state vs. two-state folding** We show that at $\phi = 0\%$, PGK unfolding and folding by pressure experiences an intermediate state, while at $\phi = 40\%$, such intermediate states vanish. We plotted the free energy profiles as a function of overlap, $\chi$, at several pressures. The overlap function $\chi$ better characterizes PGK structures from different phases than Q. Fig. S5.2(a) shows that in the absence of crowding ($\phi = 0\%$), PGK denatures by pressure from a basin of folded state at $\chi = 0$ to a basin of unfolded state at $\chi = 0.9$ through an intermediate ($\chi$ ~0.4). However, folding/unfolding is two-state in the presence of crowding (b, $\phi$ = 40%). In Fig S5.2(b), there are effectively no barriers between the Sph ($\chi$ = 0.3) and SU ($\chi$ = 0.6) states.

**Desolvation vs. macromolecular crowding** We showed that the competition between desolvation and excluded volume effect from macromolecular crowding accounts for the shift from a three-state folding mechanism to a two-state one under pressure denaturation at a denaturing condition of $\sigma^3P/\varepsilon$ = 6.6 and $k_BT/\varepsilon$ = 0.97. In Fig S5.2, we plotted a 2D free energy landscape with *pseudo Q* (the fraction of water-mediated native contact formation) and the overlap function ($\chi$) to characterize the role of water-mediated native contacts in the folding mechanism PGK in the absence (Fig S5.2a) and in the presence of crowders (Fig S5.2b). At $\phi$ = 0%, *pseudo Q* is less than 0.1 through all $\chi$ values, indicating that there are few water-mediated native contacts in the folded, intermediate, and unfolded ensembles. In the presence of crowding, the *pseudo Q* increases to 0.2 at $\chi$ = 0.3, indicating an increase in the contact of water mediated native contacts for the Sph state. At $\chi$ = 0.6, *pseudo Q* is around 0.35 (reaching a maximum of 0.5), indicating a third of the SU states are water-mediated contacts. In the presence of crowding, unfolded conformations are compact and up to half of the contacts are water-mediated contacts (Swollen Unfolded states SU). Water-mediated native contacts in the SU states form a “wet interface” between the folded and fully unfolded part of PGK. Folding does not go through collapsing a polymeric chain because the conformation is compact to begin with. Instead, folding/unfolding transition under crowding is achieved by expelling water molecules out of this wet interface PGK in the SU state to return to a compact folded state (CC or Sph). This process does not require a significant reduction in configurational entropy. The effect of macromolecular crowding flattens the energy landscape by increasing the number of water-mediated native contacts in the PGK conformations. This again shows that there is essentially no barrier between the Sph ($\chi$ = 0.3) and SU ($\chi$ = 0.6) states (Fig S5.2b).

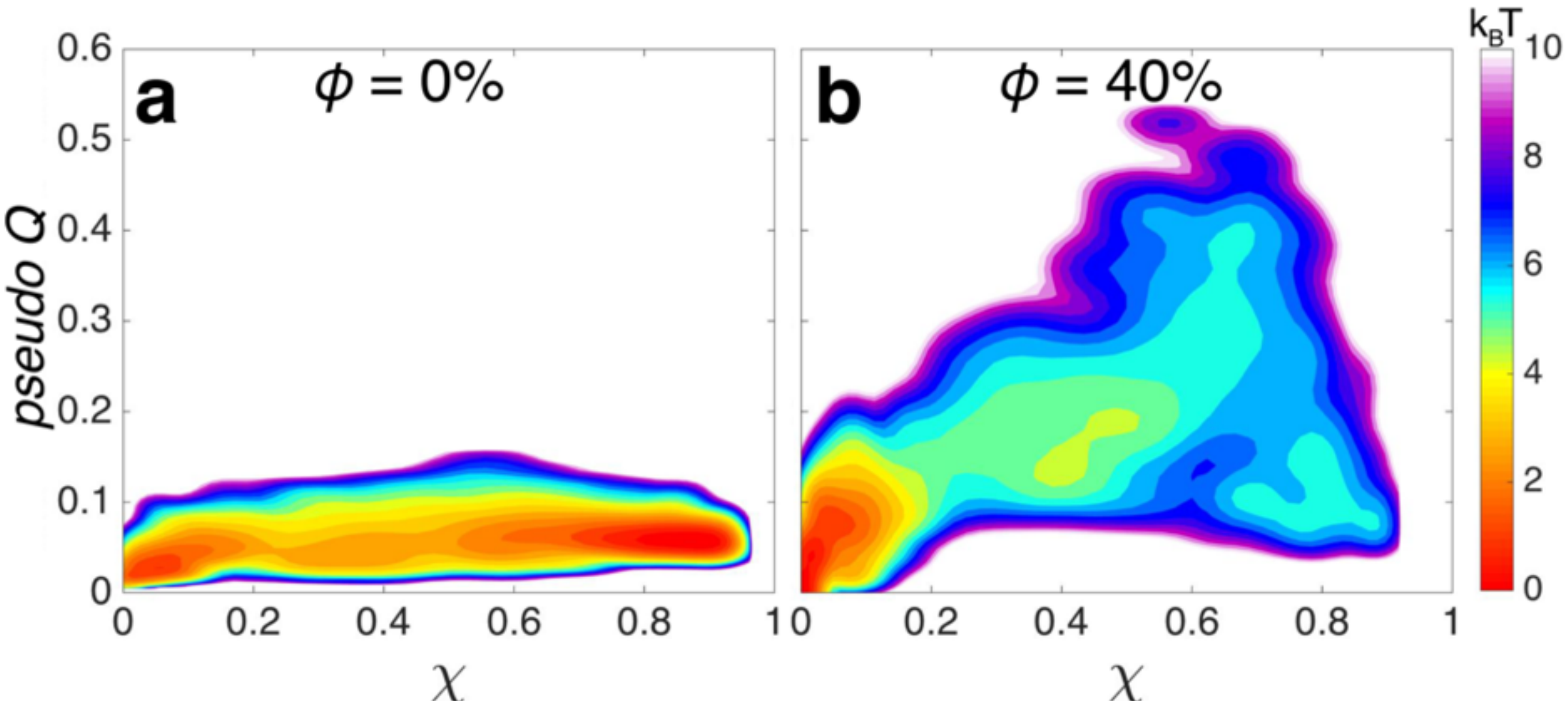

**Figure S5.2** 2D free energy diagram as a function of the fraction of water mediated contacts (*pseudo Q*) vs. overlap, ($\chi$), for PGK at $\sigma^3P/\varepsilon$ = 6.6 and $k_BT/\varepsilon$ = 0.97 without (a) and with (b) crowders.

**Evolution of PGK unfolding by pressure in the absence and presence of crowding** For PGK in the absence and presence of crowders at $\sigma^3P/\varepsilon$ = 6.6 and $k_BT/\varepsilon$ = 0.97, Figures S5.4 and S5.5 show the evolution of the probability of native contact formation, $Q$, between secondary structures, and the probability of water-mediated contact formation, *pseudo Q*, between secondary structures, respectively. In the absence of crowding, the main entry point for water is around the *m* and *n* anti-parallel beta stands (pointed out by the arrows on top of Fig. S5.4; see section S1 for structural description and schematic diagram). This region is the least stable part of the protein and causes the unfolding of the N-domain. In contrast, crowding partly stabilizes these contacts (bottom of Fig. S5.4).

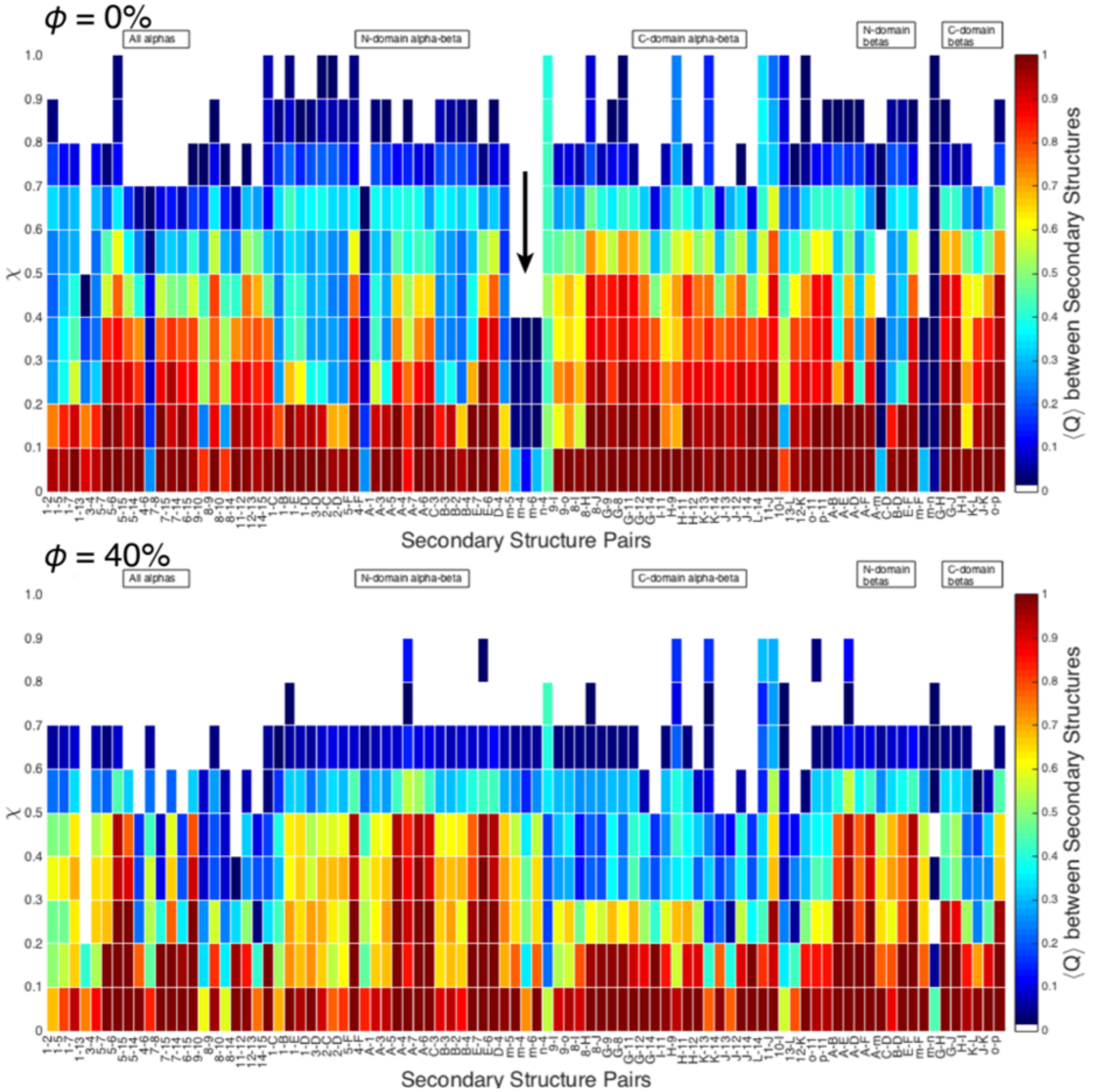

**Figure S5.4** Evolution of native contacts formation between secondary structures from $\chi$ = 0 to 1.0 for PGK at $\sigma^3 P/\varepsilon$ = 6.6 and $k_B T/\varepsilon$ = 0.97 with $\phi$ = 0% on top and $\phi$ = 40% on bottom. Numerical labels *1* through *15* are alpha helices, N-domain parallel beta-strands are letters *A* through *F*, C-domain parallel beta-strands are *G* through *L*, and *m* through *p* are anti-parallel beta strands (see table S1.1).

In the presence of crowding, the cracking primarily occurs between the alpha-helix linkers and the C-domain (between alpha-helix *7* and *8*, and between alpha-helix *14* and beta-strand *L*; these are the red regions on bottom of figure S5.5). The solvated, disordered hinge facilities conformational changes from Sph state to CC state by flexible twisting. This cracking also opens a vulnerable entry point for water to penetrate the wet core of the C-domain, causing it to unfold before the N-domain. The wet core of the C-domain is between alpha helices *9* & *10* and parallel beta strands *H* & *I* (bottom of Fig. S5.5). Here, the late "cracking" of the C-domain and increase in rigidity of the of the *m* and *n* beta stands may result in both domains unfolding cooperatively, resulting in a two-state folding mechanism.

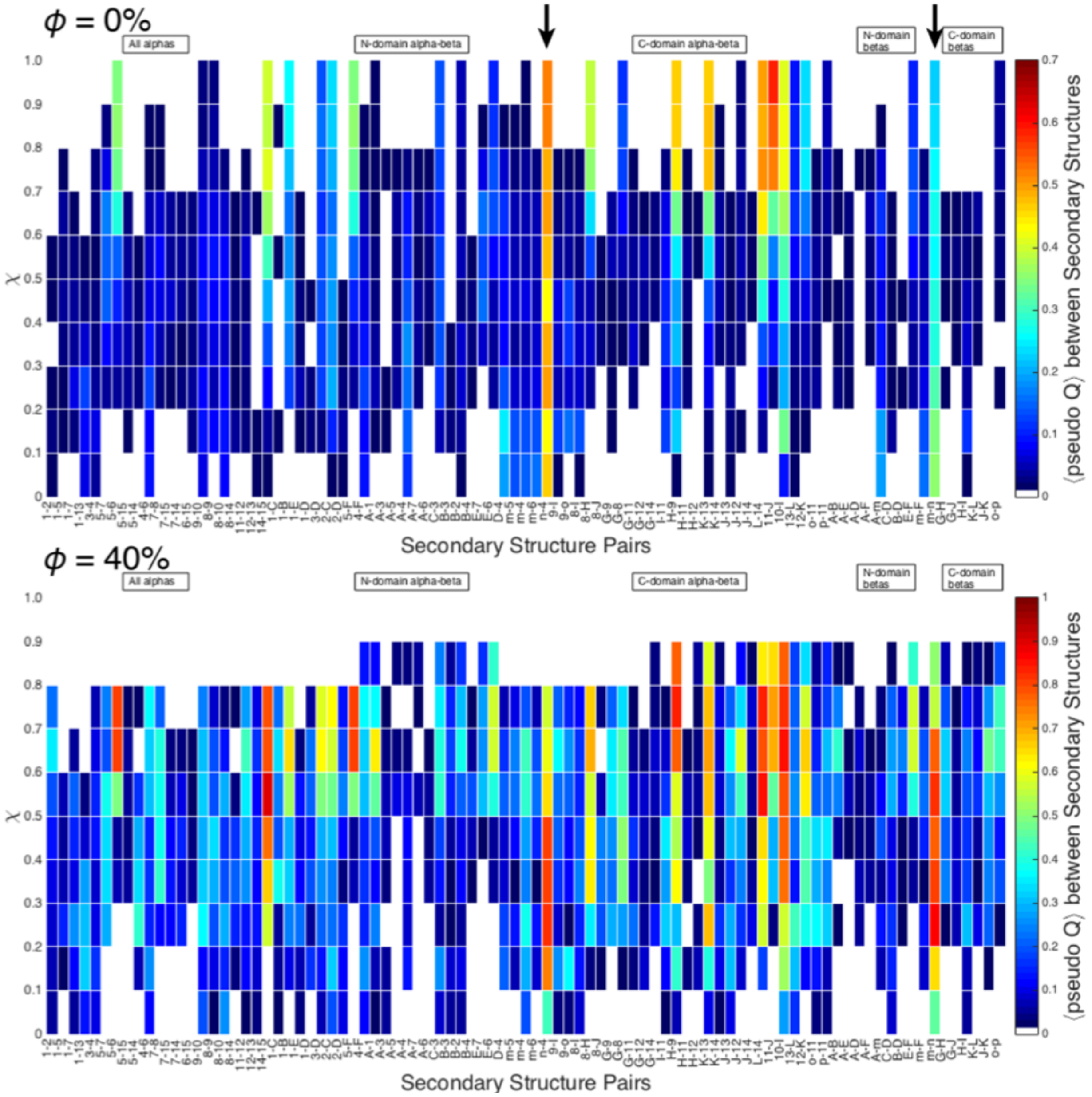

**Figure S5.5** Evolution of water-mediated native contacts formation between secondary structures from $\chi = 0$ to 1.0 for PGK at $\sigma^3 P/\varepsilon = 6.6$ and $k_B T/\varepsilon = 0.97$ with $\phi = 0\%$ on top and $\phi = 40\%$ on bottom. Numerical labels *1* through *15* are alpha helices, N-domain parallel beta-strands are letters *A* through *F*, C-domain parallel beta-strands are *G* through *L*, and *m* through *p* are anti-parallel beta strands (see table S1.1).

Lastly, the SU state can have up to half of all the native contacts be swollen with water. These water-mediated contacts expand from the initial cracking areas found in the Sph and CC state into areas in between the alpha-helices and parallel beta-strands of both domains.

**Pressure denaturation vs Thermal denaturation** We compared the folding route of PGK from pressure denaturation and those from thermal denaturation. Our high temperature ($k_BT/\varepsilon = 1.31$) simulations show that the C-domain unfolds before the N-domain (Fig. S5.3a). This agrees with experimental findings [28]. Surprisingly, at the equivalent of room temperature ($k_BT/\varepsilon = 0.97$), the pressure-dependent intermediate contains a mostly folded C-terminal domain and unfolded N-terminus domain, even though the C-terminal domain is less thermally stable than the N-domain (Fig. S5.3b). Water is more likely to penetrate the hydrophobic core of the N-terminal domain first, which triggers an unfolding sequence of the individual domains contrasting that of thermal denaturation where the C-terminal domain unfolds first. Interestingly, in the presence of crowding, the gap between the two folding route curves is closer together than in the other two plots (Fig. S5.3a and S5.3b).

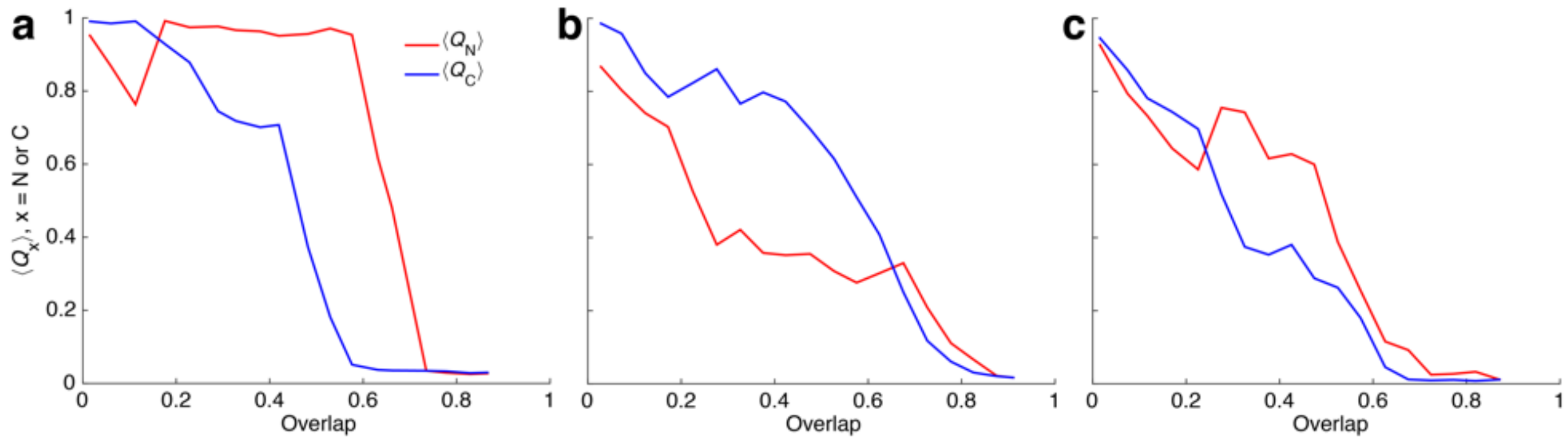

**Figure S5.3** Folding route analysis of PGK. (a) at $\sigma^3P/\varepsilon = 10^{-3}$, $k_BT/\varepsilon = 1.31$, and $\phi = 0\%$, (b) at $\sigma^3P/\varepsilon = 0.6$, $k_BT/\varepsilon = 0.97$, and $\phi = 0\%$, and (c) at $\sigma^3P/\varepsilon = 0.6$, $k_BT/\varepsilon = 0.97$, and $\phi = 40\%$ . Average local order parameters are plotted against overlap function to show their behavior as the protein folds. $Q_x$ (where $x = N$ *or* $C$) measures fraction of native contacts for the C-domain (blue) or N-domain (red). To obtain a mean field view of unfolding/folding paths, the average local order parameters $\langle Q_x \rangle$ (where $x = N$ *or* $C$) is compared with the global order parameter, overlap function. This folding route analysis uncovers an average route of folding for the domain.

## S6. Derivation of critical line

This section expands upon the derivation of the critical line from Eq. (3) and explanation in Appendix C. In order to arrive at the scaling relationship from Eq. (5) in Appendix C, in subsection S6.1, we will briefly go through the polymer model used in Ref. [30], and then calculate the crowding-dependent mean-square end-to-end distance $\langle R^2 \rangle$ in subsection S6.2. Lastly, in subsection S6.3, we derive the crowding-dependent critical temperature $T_c(\phi)$ using the calculations from the previous subsection.

### S6.1 The polymer model in a crowded solution

The Hamiltonian for the isolated polymer in a crowded solution is formulated to be,

$$\mathcal{H}[r(s)] = \frac{3}{2l}\int_0^L \left(\frac{dr}{ds}\right)^2 ds + \omega\int_0^L ds\int_0^L ds'\delta\big(r(s)-r(s')\big) + \sum_{i=1}^{N}\int_0^L V[r(s)-R_i]ds. \quad \text{(S13)}$$

The model uses a continuous curve $r(s)$, parametrized by the variable $s$, to describe the conformation of the polymer of length $L$. The strength of the excluded-volume interaction is controlled by the parameter $\omega$. The last term is the crowder potential and is given by,

$$V(r-R_i) = \beta V_0 \delta(r-R_i)/l, \quad \text{(S14)}$$

and is divided by the Kuhn length $l$.

For this model without crowders, which is the Edwards polymer model [31], the mean-squared end-to-end distance $\langle R^2\rangle$ scales by [32],

$$\langle R^2\rangle \sim L^{2v}, \quad \text{(S15)}$$

where the exponent $v = 3/5$ for $\omega \neq 0$. Whereas, $\langle R^2\rangle$ with crowders is given by,

$$\langle R^2\rangle = \frac{1}{Z(\omega,\phi,L)}\int \mathcal{D}r(s)|r(L)-r(0)|^2 e^{-S[r(s)]}, \quad \text{(S16)}$$

With $Z(\omega,\phi,L)$ being the partition function and $S[r(s)]$ as the effective action for the polymer Hamiltonian in crowded solution, Eq. (S13), with volume fraction $\phi$.

### S6.2 Self-consistent equation to find the end-to-end distance

In order to evaluate the above path integral in Eq. (S16), Ref. [30], employs a self-consistent variational approach introduced by Edwards and Singh [33]. The key reasoning behind this approach comes from choosing an effective reference action with an appropriately renormalized step length $l_1$ such that $\langle R^2\rangle \equiv Ll_1$. This ensure that all correction terms to the relation be zero, by definition. The evaluation of the path integral in Eq. (S16) results in a self-consistent equation for $l_1$ as a function of $l$, $\omega$, $\phi$, and $L$ (Appendix of Ref. [30]):

$$Ll_1^2\left(\frac{1}{l}-\frac{1}{l_1}\right) = 2c_1(1-c_0\phi)\frac{L^{3/2}}{l_1^{1/2}}, \quad \text{(S17)}$$

where $c_1 = 2\omega\sqrt{6/\pi^3}$ and $c_0 = \frac{1}{\omega}\left(\frac{\beta V_0}{l}\right)^2$. Therefore,

$$l_1^{5/2}\left(\frac{1}{l}-\frac{1}{l_1}\right) = 2c_1(1-c_0\phi)L^{1/2}, \quad \text{(S18)}$$

resulting in the scaling relation,

$$l_1 \sim (1-c_0\phi)^{2/5}L^{1/5}. \quad \text{(S19)}$$

When substituted Eq. (S19) into $\langle R^2\rangle = Ll_1$, it becomes a $\phi$-dependent version of the well-known Flory scaling relation,

$$\langle R^2\rangle \sim (1-c_0\phi)^{2/5}L^{6/5}. \quad \text{(S20)}$$

In case without crowders, $\langle R^2\rangle \sim L^{6/5}$; therefore, the ratio between with and without crowders becomes,

$$\frac{\langle R^2\rangle(\phi)}{\langle R^2\rangle(0)} \sim (1-c_0\phi)^{2/5}. \quad \text{(S21)}$$

The critical volume fraction then is [30],

$$\phi_c = \omega\left(\frac{l}{\beta V_0}\right)^2 = \frac{1}{c_0}. \quad \text{(S22)}$$

Since $\langle R^2\rangle \sim \langle R_g^2\rangle$ (where $R_g$ is the radius of gyration), the relation in Eq. (S21) also holds for $R_g$, resulting in Eq. (5) of Appendix C.

**S6.3 Crowding-dependent critical temperature**

We derived the critical line [Eq. (2) & (3); red line in Fig 4(c)] on the $T$-$P$-$\phi$ phase diagram using a simple statistical mechanical model. To begin, the Landau-Ginsberg free energy,

$$F = -r(T,\phi)\Psi^2 + u\Psi^4 + F_0, \tag{S23}$$

is used to describe the critical transition, where $\Psi$ is the order parameter, which is a scaled and shifted $R_g$ so that $\Psi = -\Psi_0$ for the I state and $\Psi = +\Psi_0$ for the U state. Since we are only interested in the critical transition, we can ignore the odd powered terms. To find the free energy minima, we take the derivative with respect to $\Psi$, and solve for the zeros of the equation:

$$\frac{\partial F}{\partial \Psi} = -2r\Psi + 4u\Psi^3 = 0$$

$$\Psi = \begin{cases} \pm\sqrt{\dfrac{r}{2u}} \\ 0 \end{cases} \tag{S24}$$

At the critical temperature $T_c$, or at critical crowding volume fraction $\phi_c$, the two phases merge together (I and U) at $\Psi = 0$, meaning $r = 0$. Furthermore, since $\Psi^2 \sim \langle R_g^2\rangle$, then,

$$\Psi^2 = \frac{r}{2u} \sim \left(1 - \frac{\phi}{\phi_c}\right)^{\gamma}. \tag{S25}$$

To find a reasonable function for $r(T,\phi)$, it must satisfy Eq. (S25) and the following limits:

$$\lim_{T\to T_c} \Psi = 0$$
$$\lim_{\phi\to\phi_c} \Psi = 0 \tag{S26}$$
$$\lim_{\phi\to 0} \Psi = \sqrt{\frac{r}{2u}}$$

Therefore, a reasonable function is

$$r(T,\phi) = -r_0\left[T - T_c^0\left(1 - \frac{\phi}{\phi_c}\right)^{\gamma}\right], \tag{S27}$$

where $T_c^0$ is the critical temperature at $\phi = 0$, and $r_0$ is a positive constant. Examples of Eq. (S23) using Eq. (S27) are plotted in Fig. S6.1. When $r = 0$, $T = T_c^0\left(1 - \frac{\phi}{\phi_c}\right)^{\gamma}$; thus, we can define a $\phi$-dependent $T_c$ as,

$$T_c(\phi) = T_c^0\left(1 - \frac{\phi}{\phi_c}\right)^{\gamma}, \tag{S28}$$

recovering Eq. (3).

To find $\gamma$, we linearly fit of Eq. (S28) on a log-log scale of the experimental $T_c(\phi)$ values. The linear fit is best when $\phi_c = 0.5$, resulting in with $\gamma = 0.40 \pm 0.01$, which is a perfect match with the exponent from Eq. (S21).

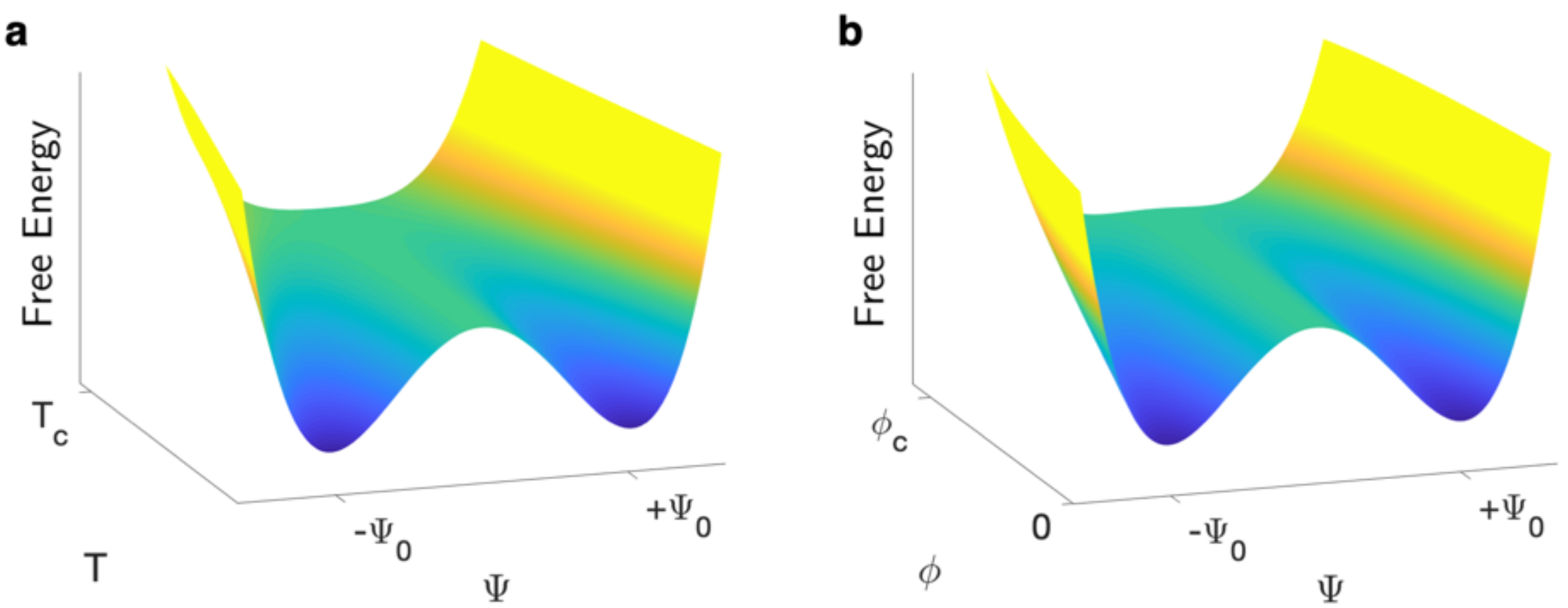

**Figure S6.1** Example Landau-Ginsberg free energies with respect to order parameter $\Psi$, and (a) $T$ with constant $\phi < \phi_c$ or (b) $\phi$ with constant $T < T_c^0$. The form of $r(T, \phi)$ from Eq. (S23) is given by Eq. (S27) with $\gamma = 2/5$. Critical transitions occur when surpassing the respected critical points, signifying the disappearance of the barrier between the two phases (I and U), and two phases become thermodynamically indistinguishable.